\documentclass[useAMS,usenatbib]{mn2e}
\usepackage{epsfig}
\usepackage{natbib}
\usepackage{lscape}
\usepackage{subfigure}
\usepackage{amssymb,amsmath}
\DeclareMathVersion{bold}

\title[20~GHz Extragalactic Imaging and Polarimetry]{Wide-Field Imaging and Polarimetry for the Biggest and Brightest in the 20\,GHz Southern Sky}

\author[Burke-Spolaor et al.]{ 
S. Burke-Spolaor,$^{1,2,6}$\thanks{Email: sburke@astro.swin.edu.au}
R. D. Ekers,$^2$
M. Massardi,$^{3,2}$
T. Murphy,$^{4,5}$
B. Partridge,$^{1}$ \newauthor
R. Ricci,$^{2,7}$
E. M. Sadler$^{4}$ \\
$^{1}$Haverford College Astronomy Department, 370 Lancaster Avenue, Haverford,
PA, 19041 USA\\
$^{2}$Australia Telescope National Facility, CSIRO, P.O. Box 76, Epping, NSW 1710, Australia\\
$^{3}$SISSA/ISAS, Via Beirut 2-4, I-34014 Trieste, Italy\\
$^{4}$ Sydney Institute for Astronomy, The University of Sydney, NSW 2006, Australia\\
$^{5}$School of Information Technologies, The University of Sydney, NSW 2006, Australia\\
$^{6}$Swinburne University of Technology Centre for Astrophysics and Supercomputing, Hawthorn VIC, Australia\\
$^{7}$Department of Physics and Astronomy, University of Calgary, 2500 University Drive NW, Calgary, AB, Canada
}

\date{In original form 2007 June 2}

\begin{document}

\maketitle

\begin{abstract}
  
  We present the wide-field imaging and polarimetry at $\nu = 20~GHz$ of seven
  of the most extended, bright ($S_{total} \geq 0.50$~Jy), high-frequency
  selected radio sources in the southern sky with declinations $\delta < -30^{\circ}$.
  Accompanying the data are brief reviews of the literature for each source.
  The results presented here aid in the statistical completeness of the
  Australia Telescope 20\,GHz Survey (AT20G) Bright Source Sample (\citealt{BSS}; Burke-Spolaor et al., in prep). The data is of crucial interest for
  future CMB missions as a collection of information about candidate
  calibrator sources. We were able to obtain data for seven of the nine
  sources identified by our selection criteria. We report that Pictor A is
  thus far the best extragalactic calibrator candidate for the Low Frequency
  Instrument (LFI) of the Planck ESA mission due to its high level of
  integrated polarized flux density ($\sim0.50\pm0.06$~Jy) on a scale of 10
  arcminutes.  Six out of the seven sources have a clearly detected compact
  radio core in our images, with either a null detection or less than 2
  percent detection of polarised emission from the nuclei. Most sources with
  detected jets have magnetic field alignments running in a longitudinal
  configuration, however PKS 1333-33 exhibits transverse fields and an
  orthogonal change in field geometry from nucleus to jets.
  
\end{abstract}

\begin{keywords}
polarization -- radio continuum: galaxies -- galaxies: magnetic fields --
galaxies: active -- cosmology: cosmic microwave background -- galaxies:
individual: Centaurus A, Centaurus B, Pictor A, PKS 0131-36, PKS 1333-33, PKS 1610-60,
PKS 2153-69, PKS 2356-61
\end{keywords}


\section{Introduction}\label{sec:intro}

The Australia Telescope 20\,GHz survey (AT20G) is a blind 20\,GHz survey of
the whole southern sky \citep{Ric04}. The project outlined in this paper
complements two sub-projects of the AT20G Survey; the first is
the AT20G Bright Source Sample \citep[BSS]{BSS}, and
the second a high-sensitivity polarisation study of a subset of the BSS (Burke et al., in prep). The collective goal of these sister projects
was to analyse the total intensity and polarisation statistical properties for
a flux density limited sample of sources selected at 20\,GHz. The BSS
used sources south of $\delta = -15^\circ$, while the Burke sample
focussed on sources south of $\delta = -30^\circ$. Both samples included all
extragalactic AT20G objects which were brighter than 0.50~Jy, and are primarily comprised of
a population of compact objects, with 85 percent showing a flat
spectrum ($\alpha>-0.5;S_{\nu}\propto\nu^{\alpha}$) at high radio frequency \citep{BSS}.

This paper enhances the two projects by providing flux density and
polarisation data for objects of angular extent beyond approximately
2.4 arcminutes, for which accurate data could not be obtained using
a single interferometric synthesis field. This paper includes sources only sources south of $\delta = -30^\circ$.


Information about nearby, bright sources is of importance to observations of
the Cosmic Microwave Background (CMB), in particular the current focus on
detecting anisotropies in the polarised CMB signature. Thompson scattering on
the Last Scattering Surface produces linear polarisation and curl-free
patterns termed ``E-modes,'' which on small angular scales provide information
about structures in the early Universe. Observations of compact CMB foreground
sources are necessary to distinguish between real E-modes and signal scattered
by gravitational lensing of the CMB on point source angular scales.
Additionally, the diffuse plasmas in bright, extended extragalactic sources
can form patterns which are termed ``B-mode'' polarisation \citep{Huetal03};
B-mode signal in the radio lobes of extragalactic sources may be unsurprising
and can be caused by complex magnetic fields in the sources and irregular
substructures of different Faraday rotation. However, the positive detection
of a similar B-mode polarisation signal in the CMB is a major goal of
increasingly sensitive measurements of CMB anisotropy. A B-mode signal in the
CMB is predicted to occur only by CMB models with tensor (gravitational wave)
fluctuations; these are in turn tied to the rapid period of expansion that may
have occured prior to the recombination epoch. The detection of B-mode CMB
anisotropies will give insight into the energy scales that occurred directly
prior to inflation and would suggest the presence of gravitational waves.
However, the amplitude of the cosmological B-mode signals is extremely small,
and will require careful attention to foreground signals that might mimic it.
On small angular scales where extended and complex foreground sources (such as
radio galaxies) are resolved, removing contamination due to extragalactic
sources becomes particularly important. For faint, confused, and unresolved
extragalactic sources in CMB data, a statistical correction may become
necessary to calibrate and interpret the results of a cosmic background B-mode
analysis.

Bright, highly polarised sources are also necessary for upcoming CMB
instruments, in particular, the ESA Planck satellite mission's low
frequency instrument (LFI), which will observe in bands between 30 and
70\,GHz \citep{BM00}. Among the source types that have been pursued as
potential flux density and polarisation calibrators for the instrument
are extragalactic sources; however, as the spectral behaviour of the
bright extragalactic population at high frequency is difficult to
model, sources selected at lower frequency are not certain to be
suitable calibrator candidates. With our observations, along with the
catalogues of \citet{BSS} and Burke et al. (in prep), we
provide a list and homogeneous observations of bright and strongly
polarised sources that are selected at a frequency closer to CMB
observing frequencies than was previously available.

In \S \ref{sec:selsum} we outline the criteria used in the selection of our
sources and summarise the literature accompanying each source. \S
\ref{sec:reduce} details the data collection, calibration, and wide-field
imaging process used to create our results. Finally, the data (full intensity
and polarisation vector images and values) are presented in \S \ref{sec:data}
with a discussion about the images, high frequency polarimetry, and the
implications of our results. \S \ref{sec:summary} provides a brief summary of
our findings.

\begin{table*}
  \centering
  {\small
  \begin{tabular}{ccccccccc}
\hline
\textbf{Source Name} & \textbf{WMAP}  & \textbf{R.A.} & \textbf{Dec.} & \textbf{redshift} &\multicolumn{2}{c}{\bf{Size (major axis)}} &\textbf{Observed Regions}   \\
\textbf{}            & \bf{ID}      &\multicolumn{2}{c}{\bf{(J2000)}} & & \emph{arcsec} & \emph{kpc} & (\emph{\# mosaic subfields})\\
 \hline
 \hline
PKS 0131-36& ---      &01:33:33.2&-36:29:11.0& 0.0298  & 365 & 206 & Observed brighter western lobe       \\
              &          &          &           &         & & & and core region (\emph{7 pointings}) \\
Fornax A      & WMAP 138 &03:22:41.7&-37:12:30  & 0.0059  & 2840 & 335 & Did not observe                      \\
              &          &          &           &         & & & (see \S\ref{sec:selsum})             \\
Pictor A      & WMAP 150 &05:19:26.0&-45:45:54  & 0.0342  & 442 & 293 & \emph{13 pointings}                  \\
              &          &          &           &         & & &                                      \\
Centaurus A   & ---      &13:25:27.0&-43:01:00  & 0.0018  & 5$^\circ$ & 648 & Observed only $\sim$12' inner        \\
              &          &          &           &         & & & double (\emph{14 pointings})         \\
PKS 1333-33   & WMAP 185 &13:36:39.0&-33:57:56  & 0.0125  & 1863 & 463 & Observed inner double (5' inner      \\
              &          &          &           &         & & & jet region) (\emph{7 pointings})     \\
Centaurus B   & ---      &13:46:49.4&-60:24:29  & 0.0129  & 960 & 239 & \emph{24 pointings}          \\
              &          &          &           &         & & &                                      \\
PKS 1610-60   & ---      &16:15:15.8&-60:39:14.0& 0.0184  & 755 & 274 & \emph{16 pointings}       \\
              &          &          &           &         & & &                                      \\
PKS 2153-69   & WMAP 190 &21:57:05.9&-69:41:23.7& 0.0283  & 110 & 61 & \emph{2 pointings}                   \\
              &          &          &           &         & & &                                      \\
PKS 2356-61   & WMAP 187 &23:58:49.0&-60:53:07  & 0.0963  & 381 & 661 & \emph{7 pointings}                   \\
              &          &          &           &         & & &                                      \\
  \end{tabular}
}
\caption{The nine sources identified by our selection criteria. Columns 1 and 2 give the common source name and the WMAP ID, respectively. Columns 3, 4 give the radio centroid position. Column 5 gives the source redshift; for redshift references, see text. Columns 6, 7 Give the angular size and calculated linear size based on a flat, $\Omega_\Lambda=0.7, \Omega_m=0.3, H_0=72$~km/s/Mpc cosmology. Column 8 gives relevant details of the observations (including number of telescope pointings for observed sources).}\label{table:mosaictable}

\end{table*} 

\section{Source selection and Literature Review}\label{sec:selsum}

The AT20G team's confirmation followup observations to their initial
transit scan survey provided a list of sources detected at 20\,GHz, from which
Burke et al. (in prep) selected a sample of sources at declinations
$-30^{\circ} > \delta$ whose total intensity exceeded 0.50~Jy. Preliminary
results of an AT20G pilot survey are available in \citet{AT20G2}; a
description of the blind scan and followup observations of the AT20G survey
are available in \citet{BSS}.
The 2.4 arcminute primary beam FWHM and the 80 to 200 metre
interferometer spacings used in the AT20G followup did not provide
accurate flux density measurements for the most extended sources, and
furthermore left cases in which sources were either: 1) Fully resolved
(and therefore undetected) by the 60 metre shortest antenna spacing of
the AT20G followup, or 2) Had compact components (hotspots, cores)
which were detected as separate, $<$0.50~Jy sources in the
followup. The Sydney University Molonglo Sky Survey (SUMSS)
\citep{Mau03} and the Parkes-MIT-NRAO (PMN) Southern
\citep{WGB94,CGW93} and Zenith Surveys \citep{WGH96} have a higher
sensitivity to extended radio sources. This is due to their more
complete Fourier coverage for short baselines, and due to the lower
observing frequency of the two surveys which allowed higher sensitivity to diffuse, steeper spectrum radio components (SUMSS at 0.843, PMN at 5\,GHz).

To recover highly resolved 20~GHz sources, we
cross-referenced multiple-component AT20G sources against
SUMSS images.  This indicated AT20G sources which appeared to be
compact components of one larger, extended object.  We then ran a
search for sources which are highly extended and thus may have been
completely resolved out by the AT20G obervations.  We expect any
extended source with $S_{20}>0.5$~Jy to have a 5~GHz flux of
$S_5>0.9$~Jy if it has a spectral index of less than -0.5, as would be
expected by a conservative estimate for the spectral index of diffuse
emission.  We thus compiled a list of the sources in the PMN catalogue
which are flagged as extended and have an integrated 5\,GHz intensity
greater than 900~mJy.  By including these objects and those selected
at 20~GHz, this sample has a nominal flux limit for extended sources
with $S_{20}>0.50$~Jy. However, due to the spatial resolution limit of PMN and our decision to include only sources flagged as extended in PMN, the sample is not complete for sources of scales between
2.4 and 4.2 arcminutes. The number of missing sources in this range is estimated to be between 4 and 10 based on source count extrapolations.

Sources which were associated with the Large and Small Magellanic
Clouds (LMC, SMC) or in the Galactic plane ($|b| < 1.5$) were excluded
from the sample, as were supernova remnants with $|b| > 1.5$.

In summation, a source was included if it:
\begin{itemize}
\item is in the region $-30^{\circ} < \delta < -70^{\circ}$
\item is extragalactic (not including LMC, SMC).
\item was detected as a multiple-component (hotspots/core) source within the AT20G followup beam with a total flux of $S_{tot}>0.50$Jy.\\
  \indent \emph{or}
\item is marked in PMN as extended, with $S_{5}>0.90$~Jy.
\end{itemize}
If a source obeyed all the above criteria, it was included in our sample.

The resulting list from the searching process consisted of nine
objects.  However, we did not observe Fornax A due to its highly
diffuse emission, most of which would remain undetected even in ATCA's
most compact configuration (furthermore, attempts to detect the
nucleus of the galaxy at 20\,GHz during unscheduled ATCA telescope
time were unsuccessful down to a few milliJansky). For some sources, only
subregions containing compact structure were observed.  The full nine
sources detected by our criteria are listed in table
\ref{table:mosaictable}, which provides information on what part of the source was observed (if
applicable).\\
\\
Below, the radio morphology of each source is described, and significant
literature and interesting properties for each object are summarised.

\indent \textbf{\emph{PKS 0131-36 (NGC612)}}, figure \ref{fig:0131}: An asymmetric, Faranoff-Reilly II
source with a bright knot in one jet at low radio frequency \citep{Mau03}.
Associated at optical wavelengths with a magnitude 15 S0-type galaxy
\citep{BGM64}. \citet{WS66} note a radio-optical morphology similar to
Centaurus A, with a dust lane crossing the core and an asymmetric positioning
of the core along the axis of the two radio lobes. Recently, a study of HI gas in the galaxy demonstrated the existence of a huge, 140~kpc disk along NGC612's dust lane and a faint bridge of HI emission to the nearby galaxy NGC619, indicating a possibly recent or ongoing interaction in this system \citep{emontsetal}.
The source resides near the centre of a cluster of galaxies
\citep{Sch75}. Redshift measured by \citet{Eke1978}.\\
\\

\indent \textbf{\emph{Fornax A (NGC1316)}}: Fornax A is a highly diffuse,
symmetrical double-lobed galaxy extending approximately one degree along its
major axis. Source is associated with the peculiar S0 galaxy NGC1316
\citep{Mills54,Ferg89}. The source resides in the Fornax cluster. The high
resolution observations of \citet{GF84} across 1.5, 4.9, and 15.0~GHz indicate
a steep-spectrum core and nonlinear jets. The large-scale structure of the
radio source was studied by \citet{Ekers83}, who noted a bridge of emission
extending between the lobes that is displaced from NGC1316's centre; this
morphological feature and the complicated dynamics of the galaxy lead to their
description of a formation scenario involving multiple galaxy collisions.
\citet{CCD94} further suggest Fornax A's host as a merger remnant, noting
tidal tails, filaments, dust lanes, and a low velocity dispersion. Although
too diffuse to make useful measurements with our telescope configuration, the
source was detected in WMAP at 23~GHz with a flux of $S_{23} = 11.7\pm 0.1$
\citep{WMAP}.  The redshift given in the table is as measured by \citet{Lon98}.\\
\\
\indent \textbf{\emph{Pictor A}}, figure \ref{fig:pica}: A classic double radio galaxy with
significant hotspots, identified both as a D type galaxy \citep{Sch75} and as
an N type galaxy with strong X-ray emission by \citet{MMB78}. It is known to
be hosted by a smooth, $e\sim0.7$ elliptical galaxy with a sharp nucleus and
little to no filamentary structure \citep{DFP77}. Extensive observations and
analysis have been done of this source. These include studies of its rotation
measure, for instance \citet{BS67} and \citet{Hav75} give $RM=46\pm 1, \theta_0=104\pm
4$ and $RM=51\pm 4, \theta_0=100\pm 6$, respectively, where $\theta_0$
is the calculated zero-wavelength polarization angle. In addition, there have been high-resolution
observations (\citet{PF83} note a radio nucleus of parsec scale) and central
black hole mass estimations (\citet{LE06} estimate $M_{BH} =
4\pm2[10^{7}M_{\sun}]$ using stellar
velocity dispersion).
The western radio lobe contains a bright hotspot that shows highly
polarised (optical polarisation degree $\sim$55 percent) optical
synchrotron emission and coincident X-ray signal from which
\citet{RM87} conclude pure synchrotron emission from radio to X-ray
wavelengths.  A deep optical study of the western hotspot by
\citet{TCM95} confirm this conclusion.  The study also notes strong
polarisation of the hotspot with a magnetic field oriented
perpendicularly to the jet axis.  A thorough X-ray study of Pictor A
was done by \citet{WYS01}, noting significant emission in the core,
bright western hotspot, and along a jet extending from the core to the
hotspot.  High quality imaging of Pictor A is available at various
wavelengths (e.g. \citet{TJR00, SSS99, MGT98, PRM97}).  Redshift
measured by \citet{Sch65}.\\
\\
\indent \textbf{\emph{Centaurus A (NGC5128)}}, figure \ref{fig:cena}: A very
complex and highly extended radio source associated with a spherical
galaxy crossed by large, absorptive dust lanes. Being the nearest and
one of the first radio galaxies identified, the literature is
incredibly extensive and too vast to summarise here. A seminal study of the large scale radio structure of the source from 500~MHz to 5~GHz was made by \citet{coop}; the optical
identification was made by \citet{BSS49}. The redshift measurement
is available from \citet{Gra78}.\\
\\
\indent \textbf{\emph{PKS 1333-33 (IC4296)}}, figure \ref{fig:1333}:
A ``triple'' radio source called thus for its bright inner jets and diffuse
outer lobes that are in alignment along the jet axes. The major lobes extend
approximately 30 arcminutes and the inner lobes are within 2 arcminutes of the
centroid \citep{GWC77,JM92}. The radio source is associated with the E1 galaxy
IC 4296, which is a companion to IC 4299 and is centrally located in the
galaxy cluster Abell 3565 \citep{YMS85,HG82}. The host galaxy is claimed to
have a dusty disk by \citet{CMZ01} and \citet{SPC02}; however, \citet{SG85}
and \citet{Mic98} saw no significant dust features. \citet{KBE86} and
\citet{KBC86} provide thorough multi-wavelength studies of the radio
polarisation and rotation measures, and studies of the optical, infrared, and
X-ray form of the source with an extensive analysis presented in \citet{KB88}.
The redshift was most recently measured by \citet{SLH00}.\\
\\
\indent \textbf{\emph{Centaurus B}}, figure \ref{fig:cenb}: This is a double-lobed source spanning
approximately 15 arcminutes, identified with a large, red E0 galaxy by
\citet{Lau77}.  The galaxy is close to the galactic plane ($b\sim 1.73$), and
as noted by \citet{JLM01}, literature flux estimates are often confused by
galactic foregrounds. However, \citet{JLM01} present a comprehensive
multi-frequency study of the radio galaxy, finding a close power-law fit to
the spectrum with a spectral index of -0.73. Extrapolating their published
measurements, we predict an integrated 20~GHz flux value of $S_{20}\sim
14.6$~Jy. The redshift quoted in the table was measured by \citet{WT89}.\\
\\
\indent \textbf{\emph{PKS 1610-60}}, figure \ref{fig:1610}: The radio source is a wide angle
head-tail galaxy with bright inner lobes and diffuse tails (mapped by
\citet{SM75,JM92,CFW77,GVS94} at 408, 843, 1415, and 4850~MHz, respectively).
\citet{Eke69a} identified the source with an E3 galaxy at J2000
$\alpha=$16:15:1.7, $\delta=-60$:55:13.7. The galaxy is thought (but not yet
proved) to be a central member of the local massive cluster Abell 3627
\citep{RR90,BB91,BNS96}. Another head-tail galaxy lies within one degree of
PKS 1610-60; \citet{JM96} mapped the two sources at 1360~MHz, explaining PKS
1610-60's morphology (sharp kinks in the jets and the wide tail separation)
through probable interaction with an intercluster medium and influence from
subcluster mergers. Most recently, a redshift was determined by
\citet{WKC04}.\\
\\
\indent \textbf{\emph{PKS 2153-69}}, figure \ref{fig:2153}: Less than 3 arcminutes in extent, the
radio structure of this Seyfert II galaxy consists of a core-lobe morphology.
A presumably unrelated source lies approximately 2.5 arcminutes to the east
\citep{Eke69b,CFW77}. The host galaxy is an early type of class E0
\citep{Sch75,MA82}, and has been shown to have an extra-nuclear ionised gas
cloud with which the jets are interacting \citep{TFD88,TJR96}. Further deep
radio and X-ray studies have been performed by \citet{FMW98} and
\citet{YWT05}. The galaxy was modelled to have a long-period ($t\sim 1.8\times
10^6$y) jet precession by \citet{LZ05}. Redshift given by \citet{DPD91}.\\
\\
\indent \textbf{\emph{PKS 2356-61}}, figure \ref{fig:2358}: A Faranoff-Riley II galaxy marked
by four bright regions of emission that are slightly asymmetric about
a core. In addition, there is a diffuse arm of emission extending
south-west from the western inner lobe (see e.g. low radio frequency
images\footnote{An unpublished image of this source is
available from the ATNF at
\mbox{http://www.narrabri.atnf.csiro.au/public/images/pks2356-61}} by
\citet{CFW77,JM92,SSH96,BH06}). Identified with an E3 galaxy by
\citet{Whi72} and \citet{Sch75}. Suspected to be a member of galaxy
cluster SC 2357-61 \citep{WS66,TCG90}. Source was detected as a hard
X-ray emitter by \citet{LMB05}. Redshift measured by \citet{LPM96}.

\section{Observations and Data Reduction}\label{sec:reduce}
All sources were observed with the Australia Telescope Compact Array
(ATCA). The compact configuration H75 has five 22m dishes extending on
perpendicular north-south and east-west arms with a maximum baseline
of approximately 80 metres. The array was well-suited to obtain
sufficient (\textit{u,v}) coverage to sample and image large, diffuse
structures at high radio frequencies over a short period of time. Each
source was observed simultaneously at 16.7 and 19.4\,GHz in two
polarisations using linear feeds with 128~MHz bandwidth at each
frequency. Observations took place on October 1, 2006.

The dish size and maximum baseline correspond to a field of view of
2.6 arcminutes and a resolution of 43 arcseconds at our mean
observing frequency of 18.0~GHz. Because all the sources extended
beyond the primary beam size of the telescope, each source was
observed using ATCA's ``mosaic mode''; the mosaicing technique is
commonly used for wide-field imaging at the ATCA. Overlapping,
adjacent telescope pointings that were later jointly deconvolved as
one observation were taken in a hexagonal pattern with field centres
spaced at approximately 1.5 arcminutes. The field spacings, given by
the Nyquist sampling limit for a hexagonal grid,
\begin{center}
\begin{equation}
\theta = \frac{\lambda}{\sqrt{3}D}
\end{equation}
\end{center}
(where $\theta$ is the angular size of pointing separation and $D$ is
the primary beam FWHM) ensure a constant noise level across
overlapping regions of observation. Additionally, the joint
processing of the hexagonally interleaved telescope pointings give the
interferometer a sensitivity to large scale emission that is comparable to
single-dish (zero-spacing) scales. Each mosaic subfield was observed
for 40 seconds, and a secondary calibrator was observed before and
after each mosaiced source. Calibrator and mosaic sets were performed
four times for each source at a range of hour angles.

All data were calibrated 
using the Miriad software package \citep{STW95}.
Throughout calibration the two IF bands were treated
independently. After opacity corrections and xy feed phase difference
solutions were applied, the bandpass was calibrated using a standard bandpass
calibrator (PKS B1921-293), and a primary flux density scaling was solved for
using a standard ATCA primary flux calibrator, PKS B1934-638. These
calibrators were both observed once only, directly prior to the observing run.
Each secondary calibrator was used to simultaneously solve for antenna gain matrices, the
residual xy-phase difference, and polarisation leakage terms (which result
from slight imperfections in feed alignment). Solutions from this process were
then applied to each subfield. In the case of Centaurus A, additional
self-calibration with a single Gaussian component model was needed for fields
containing the bright core of the source.

Due to current limitations and software incompatibility in some of the deconvolution algorithms, our measurement of total polarised flux and the images published in this work show the results of Miriad's joint deconvolution polarisation maximum entropy method algorithm ({\sc pmosmem}), while the total intensity measurements are extracted from a new multi-scale clean deconvolution algorithm implemented in the Common Astronomical Software Applications (CASA) software package \citep{cornwellmultiscale,CASA}.
In Miriad's joint deconvolution scheme,
all subfields are inverted as one image, and the dirty Stokes Q and U maps are
set to the same gridding as the total intensity image. The algorithm performs a maximum
entropy method deconvolution simultaneously for all polarisations, using
morphological information from bright emission in the total intensity image to solve for the
Q and U polarised Stokes images. Images are then restored using a gaussian
with the same half-power beam width of the synthesised beam\footnote{See http://www.atnf.csiro.au/computing/software/miriad for further details of mosaic observations}. While this method does produce a morphologically correct total intensity image, the MEM algorithm is not optimum for computing total intensities, causing integrated intensity measurements to be systematically high by about 1 to 20 percent. For this reason, we used CASA's multi-scale clean (again joining mosaic subfields) to deconvolve images and gather intensity information. While CASA does compute polarisation values, a known error in the data interface for ATCA polarisation information currently renders their output erroneous.

Polarisation images were created by combining aligned Q and U images
pixel by pixel, calculating total polarised intensity, $P$, using $P =
\sqrt{Q^2 + U^2}$. A noise term, $\sigma_p$ was estimated from the RMS
of a region of the Stokes Q image that did not contain source
emission. Polarised signal below 3$\sigma_p$ was considered undetected
and was masked from the polarised intensity map. The masked polarised
maps were then corrected for a debiasing factor\footnote{The debiasing
term is necessary to correct the Ricean bias that occurs from the
calculation of P from Q and U terms; there are multiple methods of
correction as detailed in \citet{vlamemo161}}, using a first order
correction that again uses $\sigma_p$: $P_{final} = \sqrt{P^2 -
\sigma_p^2}$.

Position angle maps are also calculated on a per pixel basis, with
$\psi = \frac{1}{2}\arctan{\frac{U}{Q}}$, where $\psi$ is the observed
angle of polarisation. Vector maps of fractional polarised intensity
were formed by dividing the masked total polarisation maps by the
total intensity maps; in this way, the maps will only show fractional
polarisation above the detection threshold. Polarisation vectors were
then calculated using lengths determined by fractional polarisation
levels, and position angles from PA maps. Note that the orientation of
a polarisation ``vector'' is ambiguous by 180 degrees, and does not
provide a preferred electric or magnetic field direction along the
printed rods.

\begin{table*}
  \centering
  {\footnotesize 
    \begin{tabular}{ccccccccc}
      \hline
      \textbf{Source Name}    & \textbf{RA (core)} & \textbf{Dec (core)} &\textbf{$S_{core}$}&\textbf{$S_{18}$}& \textbf{$P_{18}$}&\textbf{$\Pi_{18}$}&\textbf{$\alpha_5^{18}$}&\textbf{$S_{23}$}\\
      \hline
      \hline
      PKS 0131-36 &01:33:57.9$\pm3.0$  & -36:29:35.3$\pm1.9$ & 0.03$\pm$0.01 &     $>$0.44            & 0.02$^*$              &4.6    & ---      &---  \\
      Pictor A                   &05:19:49.7$\pm0.8$  & -45:46:43.7$\pm0.5$ & 1.32$\pm$0.04 &     6.32$\pm$0.11&0.50$\pm$0.06 &7.9  & -0.70 &6.80 \\ 
      Centaurus A          &13:25:27.6$\pm0.7$  & -43:01:04.9$\pm0.4$ & 5.98$\pm$0.17 &     $>$28.35          & 3.81$^*$              &13.4  & ---      &46.2$^{\dag}$\\ 
      PKS 1333-33        &13:36:39.0$\pm4.1$  & -33:57:57.7$\pm2.6$ & 0.30$\pm$0.05 &     $>$0.74             & 0.07$^*$              &9.8   & ---      &1.70 \\
      Centaurus B          &13:46:49.1$\pm0.3$  & -60:24:30.0$\pm0.2$ & 5.02$\pm$0.06 &     8.89$\pm0.43$ &0.08$\pm$0.01 &0.9   & -0.87 &---\\ 
      PKS 1610-60        &16:15:05.6$\pm8.8$  & -60:54:27.1$\pm8.9$& 0.14$\pm$0.05  &     2.11$\pm$0.04 & 0.128$\pm$0.008 &6.0   & -0.93 &1.7$^{\dag}$ \\
      PKS 2153-69        &   ---                               &    ---                               &    ---                      &     3.40$\pm$0.21 & 0.05$\pm$0.03 &1.5   & -0.96 &3.60 \\ 
      PKS 2356-61        &23:59:04.9$\pm7.8$  & -60:55:03.4$\pm6.1$& 0.09$\pm$0.03  &     1.64$\pm$0.05 & 0.032$\pm$0.004 &1.9   & -0.94 &1.80 \\ 
    \end{tabular}
  }  
\caption{See section \ref{sec:resimgdes} for more detailed column descriptions: (1) Common source name (2-3) J2000 RA/Dec (4,5) Core and total integrated flux density and error in Jy (6) Integrated polarised intensity and error in Jy. $^*$ indicates the integrated polarisation over an observed subregion (7) Fractional polarisation ($100\cdot P_{18}/S_{18}$) (8) 4.85-18GHz spectral index (9) Flux density as detected in WMAP maps, with \dag marking fluxes taken from the NEWPS catalogue \citep{newps}. Note that the Centaurus A measurements can be taken as actual values (not simply limits) when referring to the inner lobes of Centaurus A.}\label{table:numbers}
\end{table*} 

\section{Images and Discussion}\label{sec:data}
\subsection{Results and Image Description}\label{sec:resimgdes}
Table \ref{table:numbers} gives various observed and calculated parameters
from the data. The columns are as follows: (1) Source name; sources with a
detected nucleus have a core position fitted with a point source model, which also provides a measurement for the flux
density of the nucleus; (2-3) J2000 position of galaxy nucleus, if
detected; (4) Core flux density and error in Jy.  (5) Total flux density and
error in Jy; sources with lower limits for total integrated flux density are
those for which we believe we have acquired a minimal measurement for the flux
density due to observing a subregion of the source. Total source fluxes
were taken from an integration over the
MSclean-deconvolved maps, normalised by the synthesised beam. Note that this process has
corrected for the primary beam. Errors were quantified
by the image RMS; (6) Integrated \emph{detected} ($>3\sigma_p$)
polarised intensity in Jy ($P=\sqrt{(\sum Q_i)^2 + (\sum U_i)^2}$, where the
sums represent the integrated and beam-normalized values over the $P >
3\sigma_p$ and Stokes $I > 3\sigma$ pixels in the deconvolved Stokes Q and U
images). Because of possible depolarisation effects when imaging the missed
source regions, we cannot provide accurate total integrated polarisation
measurements for sources which were incompletely observed. Such
measurements, indicated with an asterisk, give the integrated value of the observed source subregion; (7) The fractional polarisation of the observed
source region in percentage (8) The spectral index between
4.850 and 18.0~GHz, using PMN flux densities for all sources except Pictor A,
where the 5~GHz measurement of \citet{Kuehr81} was used and Centaurus B, for which the 5~GHz measurement of \citet{pkscat} was used ; (9) WMAP flux density
at 23~GHz \citep{WMAP}; flux densities marked with \dag~are as given in
the NEWPS catalogue \citep{newps}.

\begin{figure*}

\caption{\textbf{PKS 0131-36:} \small The beam shape is indicated in the lower right hand corner of the
images in Panels (a) and (b); \textbf{\emph{(a)}} Radio reference map and greyscale image of total intensity.  An "X"
indicates the position of the associated optical galaxy.  The reference radio
map was constructed using an 843~MHz image from the SUMSS postage stamp
server, available at \mbox{http://www.astrop.physics.usyd.edu.au/SUMSS}, while the greyscale image was created from our 18~GHz mosaic data.
Note that the apparent termination of the source at the edges of the
observation field is due only to decreased sensitivity near the edge of our field of view. \textbf{\emph{(b)}} 18~GHz polarisation map; vectors show the \emph{observed} degree and
orientation of the electric field vectors, while contours trace the total
intensity of the radio source at levels given in the text below the image.
The rod in the lower right hand corner of the image indicates the length of a
100 percent polarised component.}\label{fig:0131}

\subfigure[]
{
 \includegraphics[width=0.7\textwidth]{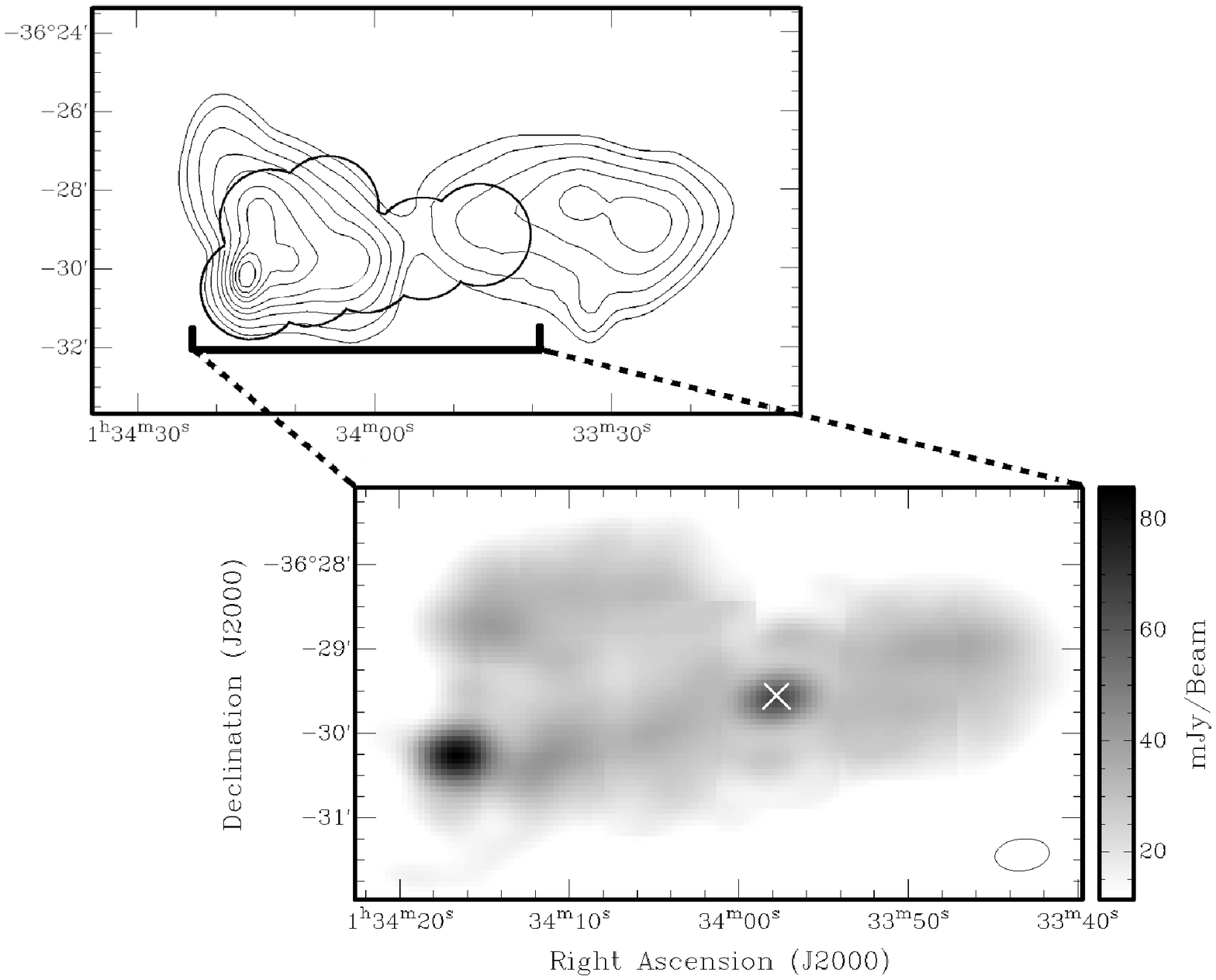}
}

\subfigure[]
{
 \includegraphics[width=0.8\textwidth]{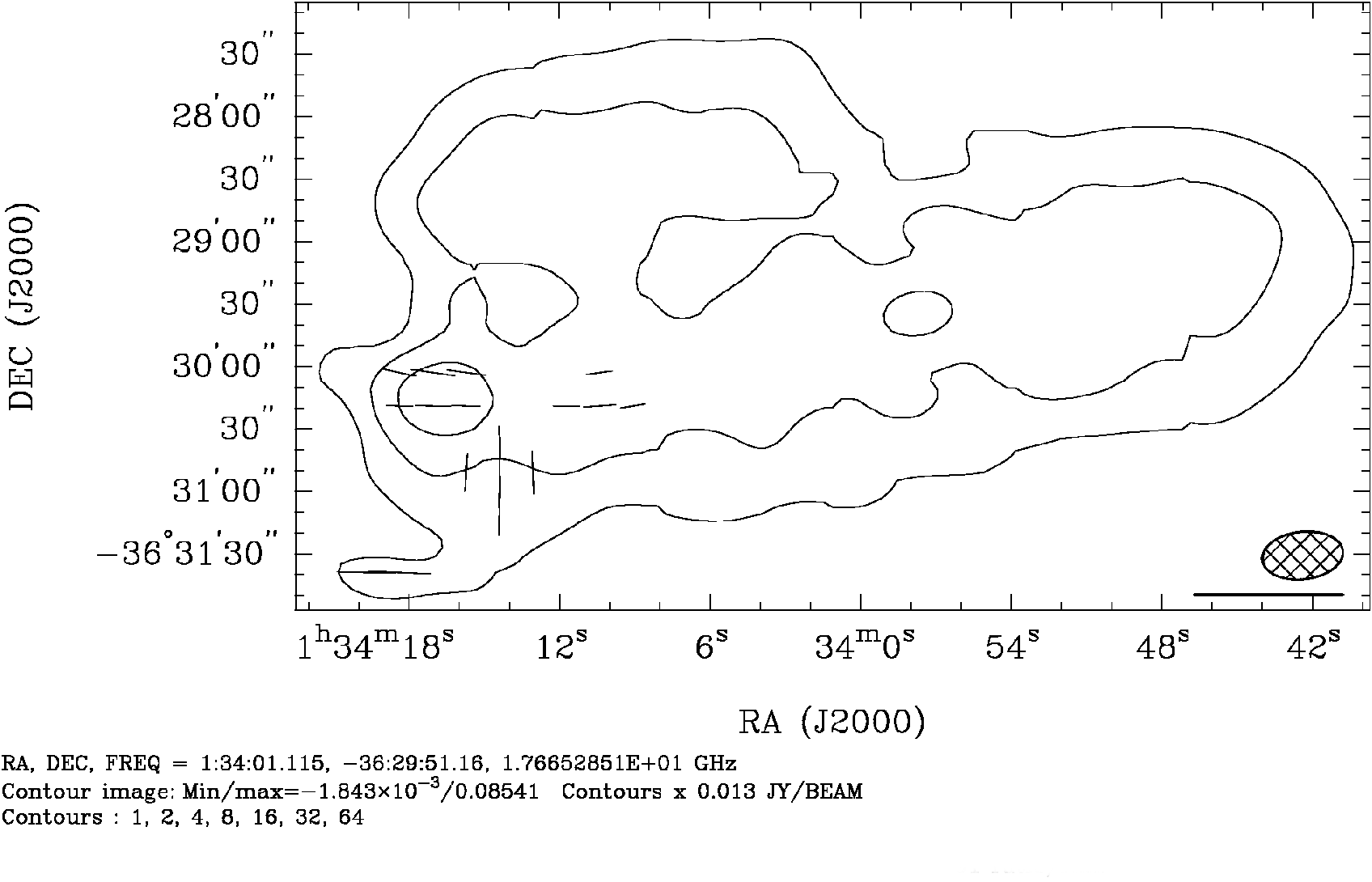}
}
\label{fig:ext}
\end{figure*}

\begin{figure*}

\caption{\textbf{Pictor A:} \small  The beam shape is indicated in the lower left hand corner of the
  images in Panels (a) and (b); \textbf{\emph{(a)}} The 18~GHz greyscale image of total intensity. An "X"
indicates the position of the associated optical galaxy. \textbf{\emph{(b)}} 18~GHz polarisation map; vectors show the \emph{observed} degree and
orientation of the electric field vectors, while contours trace the total
intensity of the radio source at levels given in the text below the image.
The rod in the lower left hand corner of the image indicates the length of a
100 percent polarised component.}\label{fig:pica}
 \subfigure[]
{
 \includegraphics[width=0.9\textwidth]{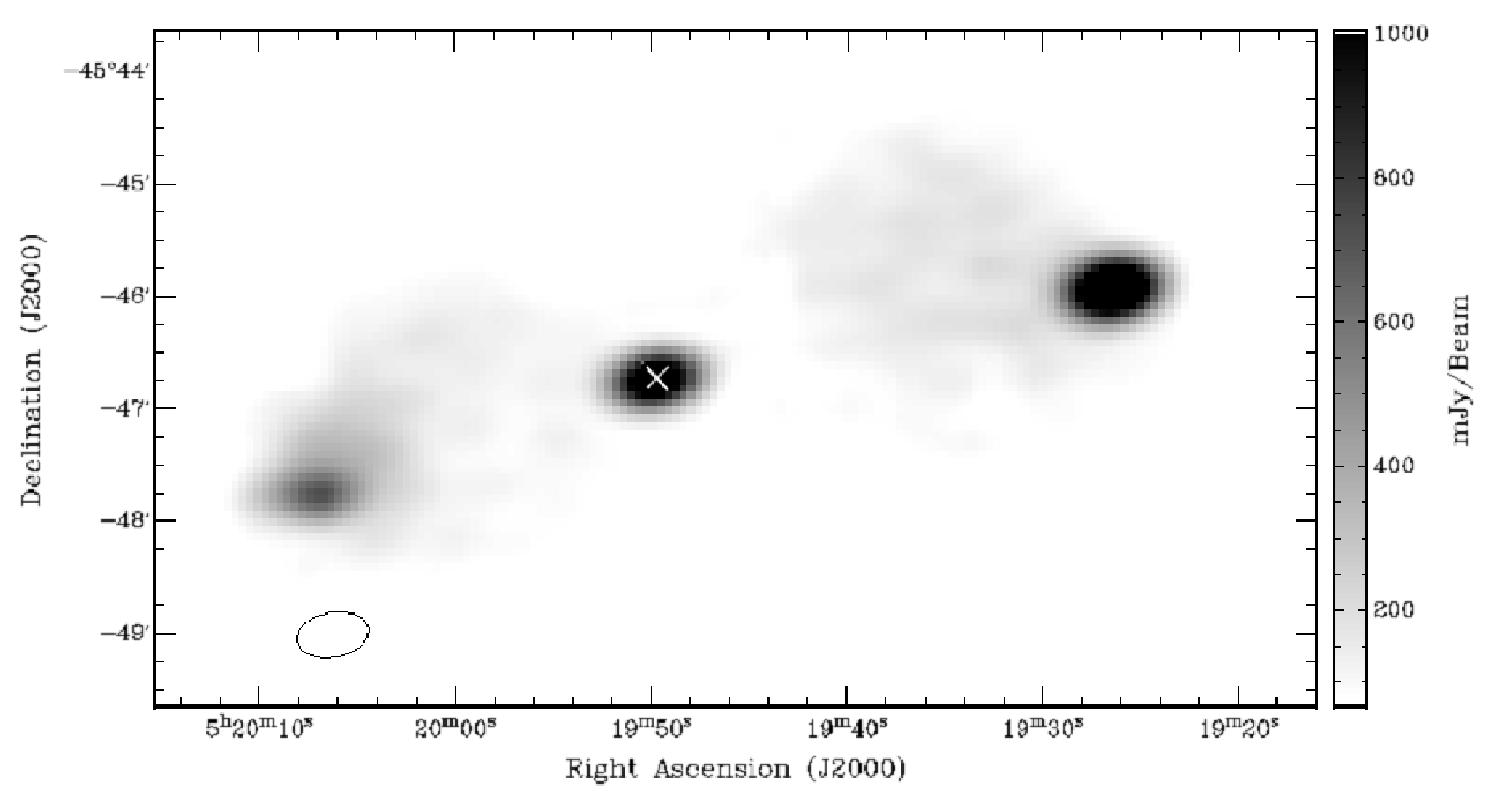}
}

\subfigure[]
{
 \includegraphics[width=0.9\textwidth]{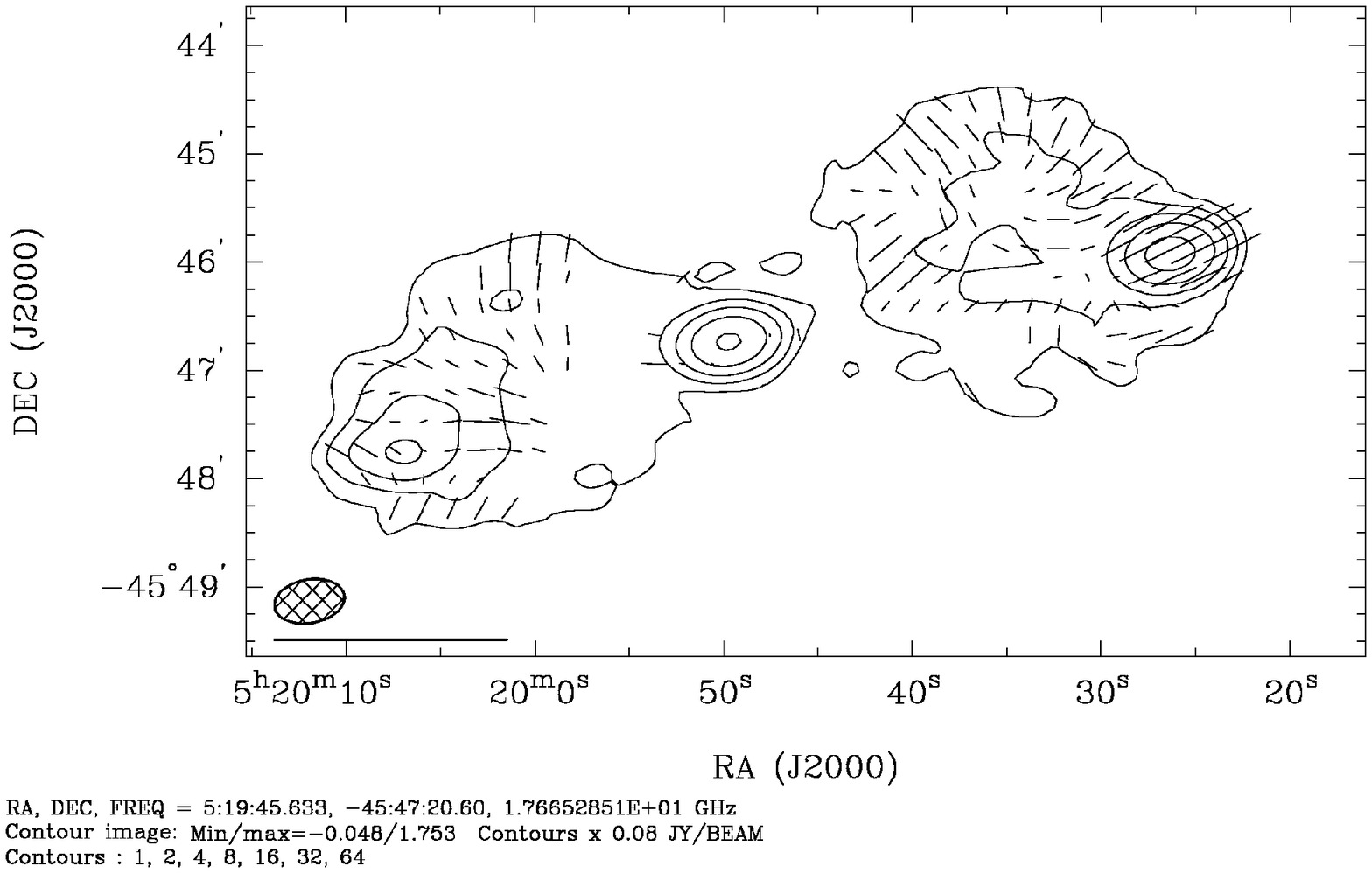}
}

\end{figure*}

\begin{figure*}

\caption{\textbf{Centaurus A:} \small The beam shape is indicated in the lower left hand corner of the
images in Panels (a) and (b); \textbf{\emph{(a)}} Radio reference map and 18~GHz greyscale image of total intensity.  An "X"
indicates the position of the associated optical galaxy. The 5.0\,GHz
reference radio map was reproduced from Ron Ekers' 1977 single-dish
observations. \textbf{\emph{(b)}} 18~GHz polarisation map; vectors show the \emph{observed} degree and
orientation of the electric field vectors, while contours trace the total
intensity of the radio source at levels given in the text below the image.
The rod in the lower left hand corner of the image indicates the length of a
100 percent polarised component.}\label{fig:cena}

\subfigure[]
{
 \includegraphics[width=0.7\textwidth]{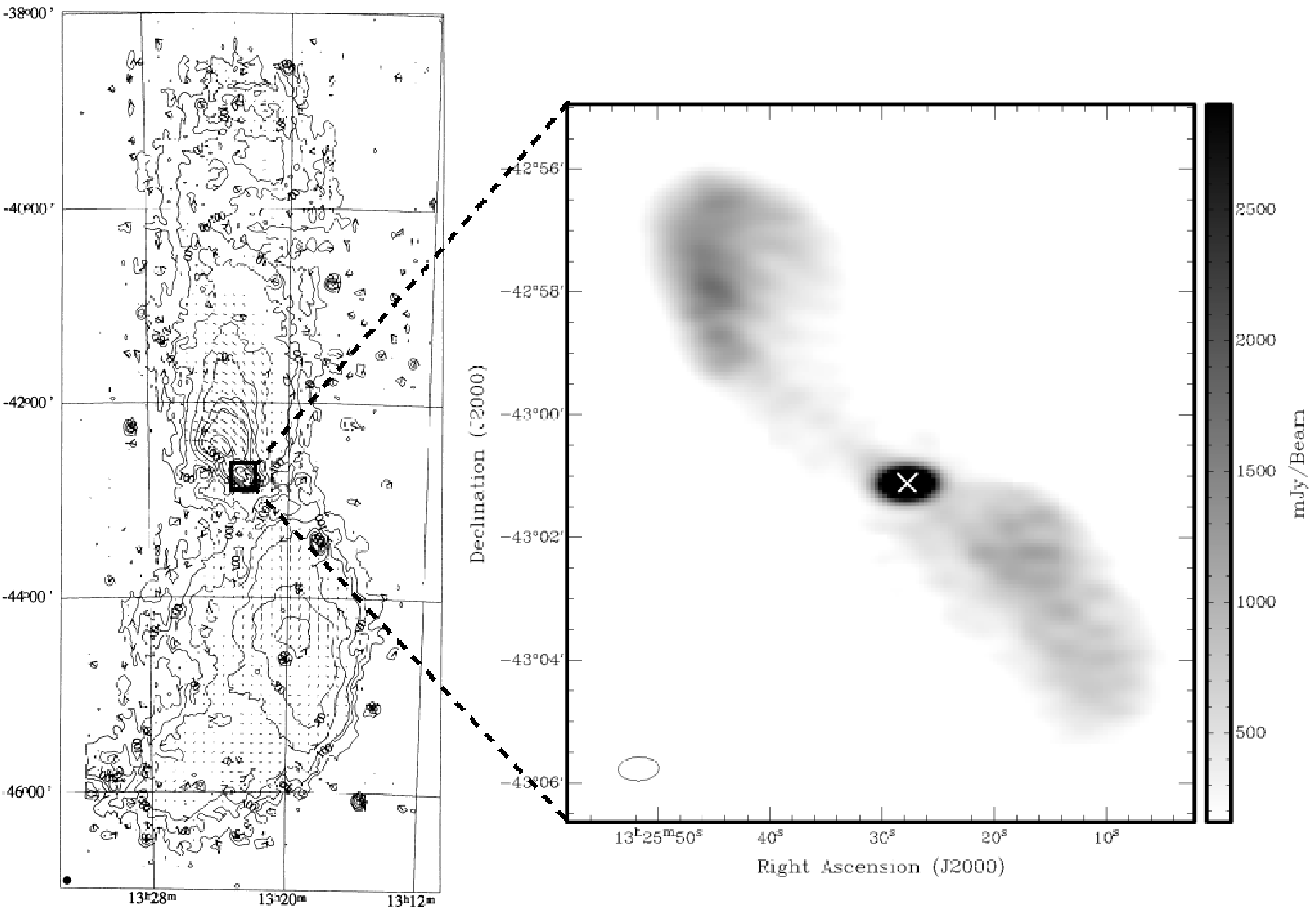}
}

\subfigure[]
{
 \includegraphics[width=0.6\textwidth]{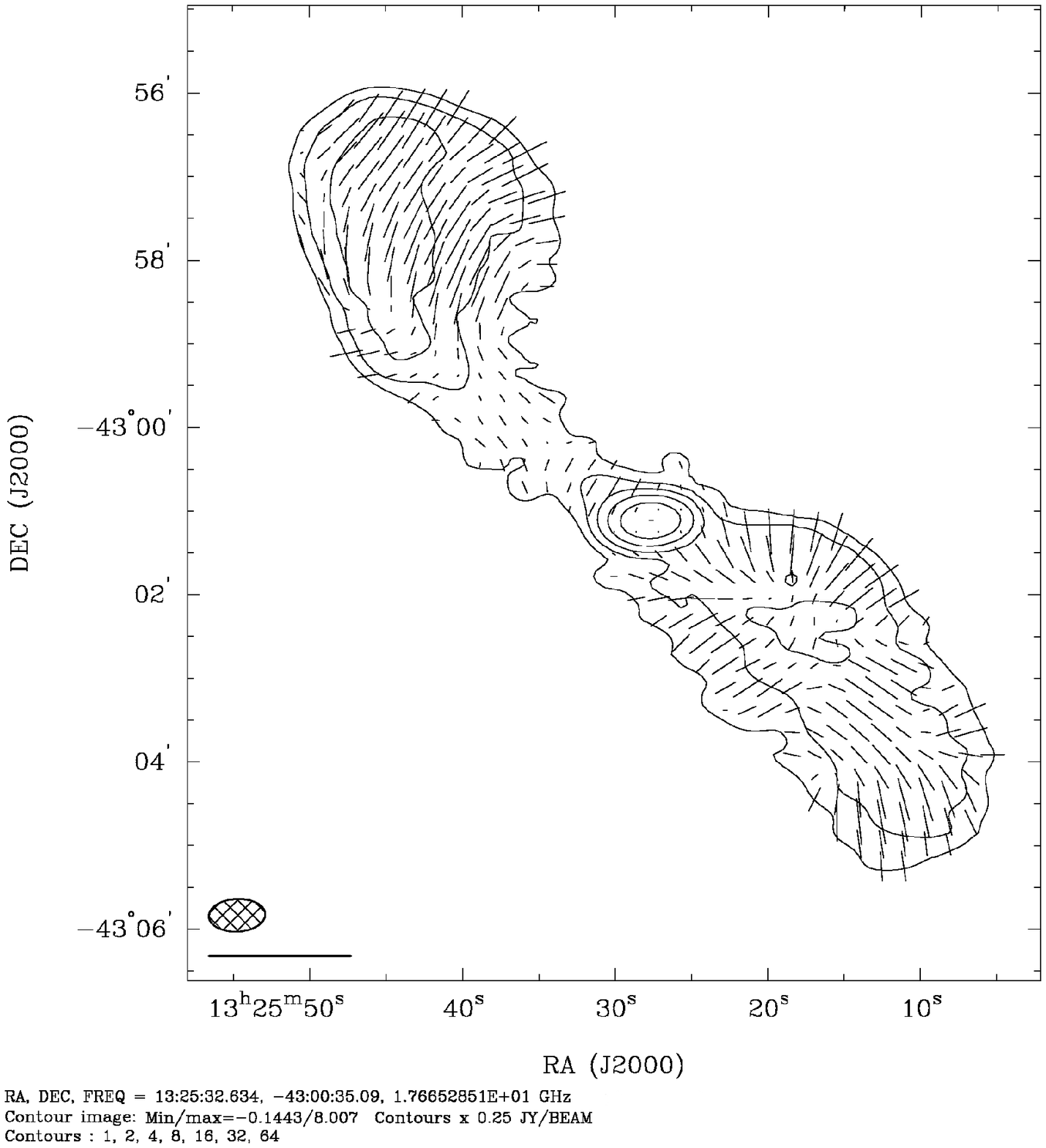}
}
\end{figure*}

\begin{figure*}

\caption{\textbf{PKS 1333-33:} \small The beam shape is indicated in the lower right hand corner of the
images in Panels (a) and (b). \textbf{\emph{(a)}} 843~MHz radio reference map reproduced from \citet{JM92} with
permission from Bruce McAdam, and 18~GHz greyscale image of total
intensity.  An ``X'' indicates the position of the associated optical
galaxy. \textbf{\emph{(b)}} 18~GHz polarisation map; vectors show the \emph{observed}
degree and orientation of the electric field vectors, while contours
trace the total intensity of the radio source at levels given in the
text below the image. The rod in the lower right hand corner of the
image indicates the length of a 100 percent polarised component.}\label{fig:1333}
 \subfigure[]
{
 \includegraphics[width=0.9\textwidth]{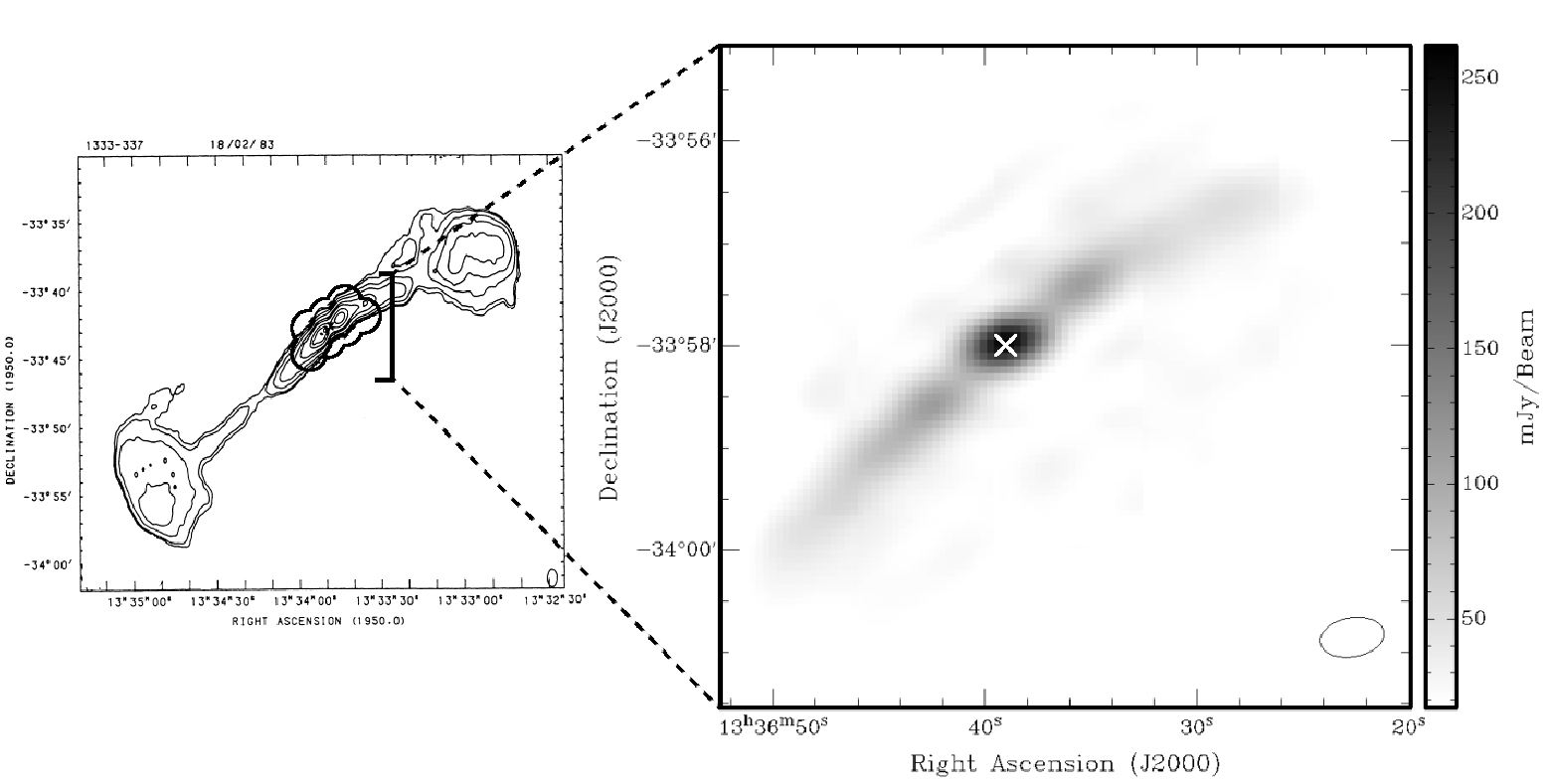}
}

 \subfigure[]
{
 \includegraphics[width=0.7\textwidth]{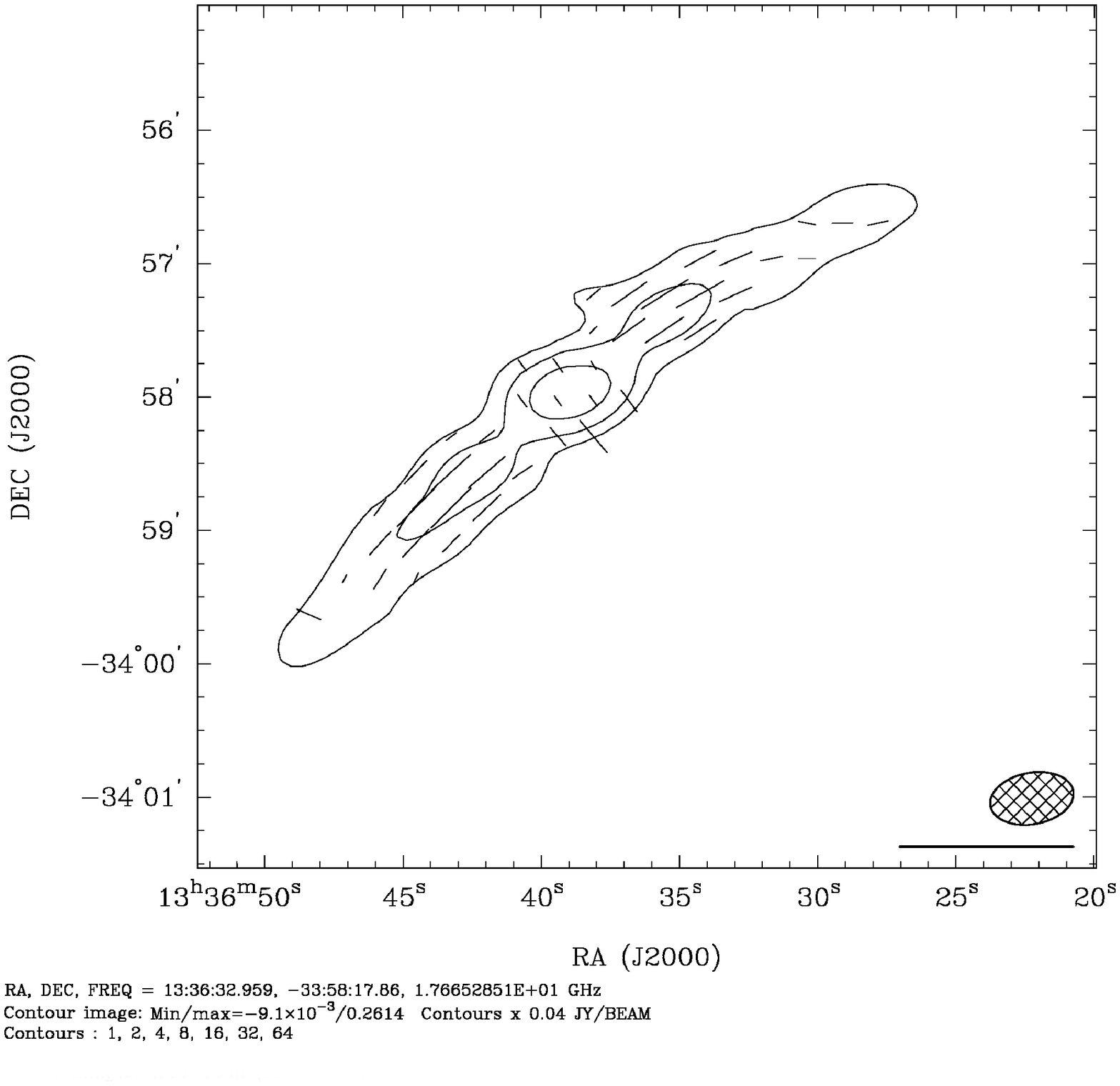}
}
\end{figure*}

\begin{figure*}

\caption{\textbf{Centaurus B:} \small The beam shape is indicated in the corner of the
  images in Panels (a) and (b). \textbf{\emph{(a)}} The 843~MHz Radio reference map for
  the observed regions covering the core, eastern lobe, and western lobe,
  reproduced from SUMSS postage stamp server. \textbf{\emph{(b)}} 18~GHz total intensity
  greyscale for the eastern lobe, core, and western lobe, respectively. An
  ``X'' indicates the position of the associated optical galaxy. \textbf{\emph{(c)}}
  18~GHz polarisation maps for the corresponding regions; vectors show the
  \emph{observed} degree and orientation of the electric field vectors, while
  contours trace the total intensity of the radio source at levels given below
  the image. The rod in the corner of each image
  indicates the length of a 100 percent polarised component.}\label{fig:cenb}
 \subfigure[] 
 {
 \includegraphics[width=0.5\textwidth]{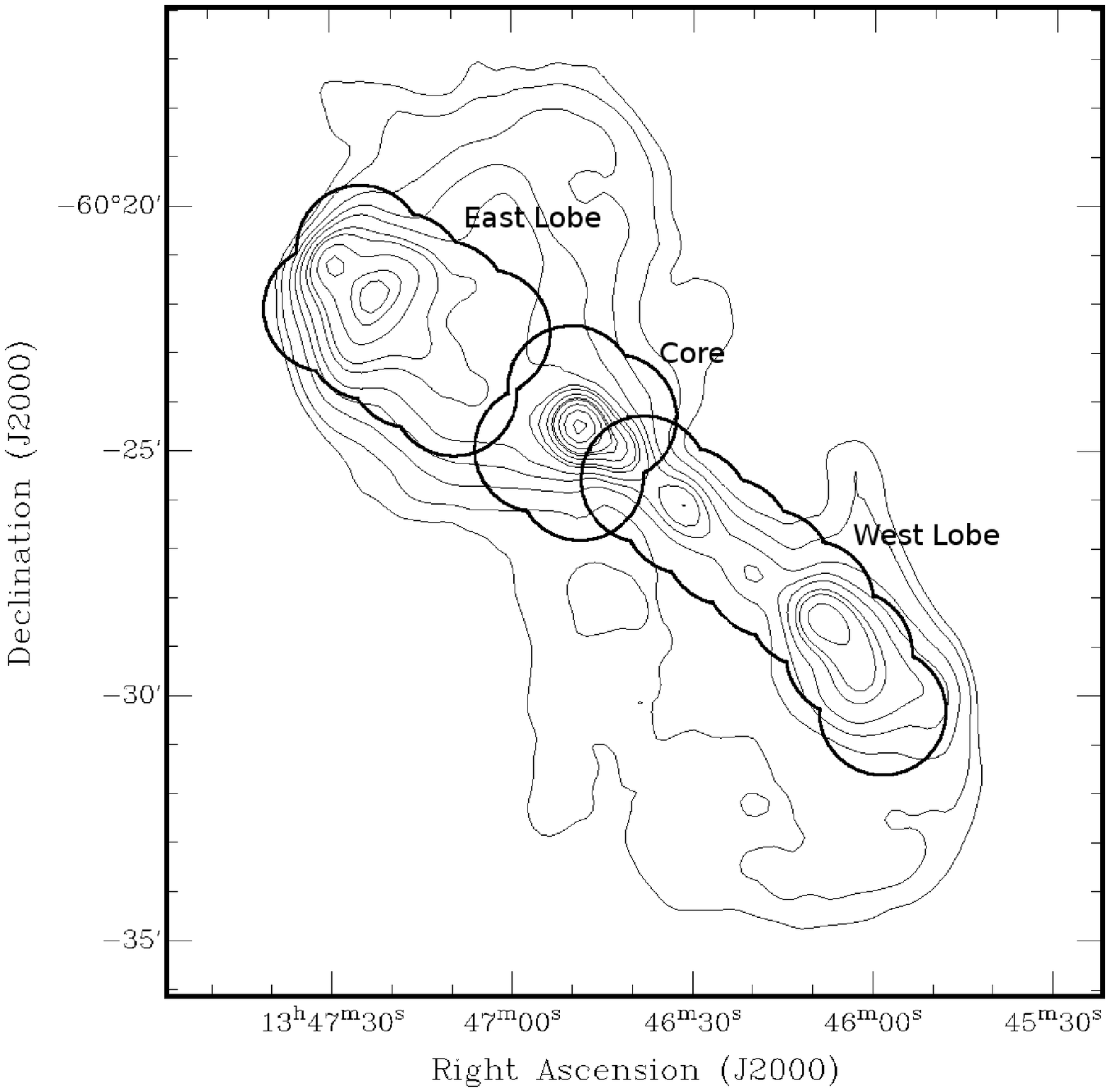}
 }
 \subfigure[]
{
 \includegraphics[height=0.31\textwidth,angle=270]{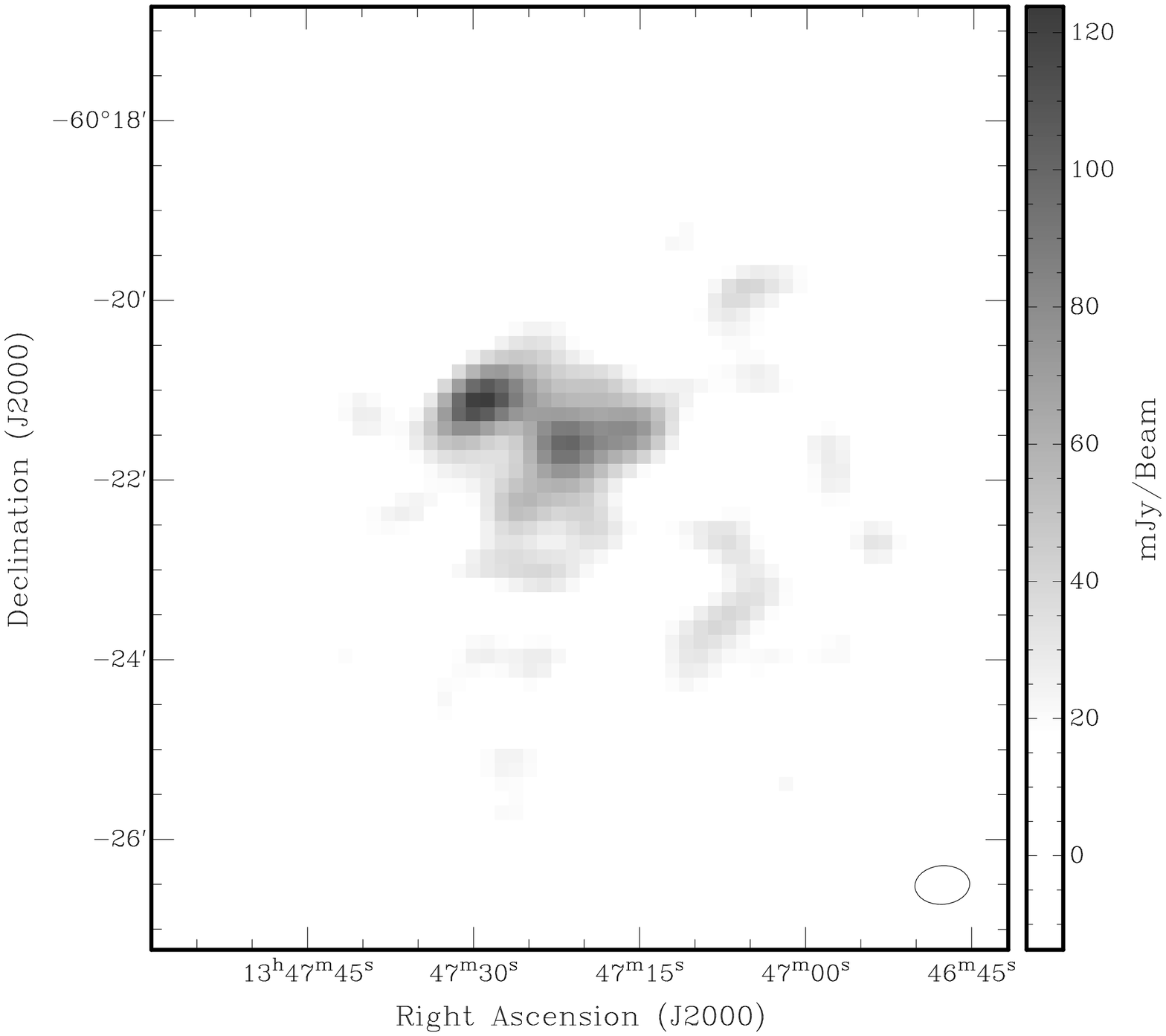}
 \includegraphics[height=0.31\textwidth,angle=270]{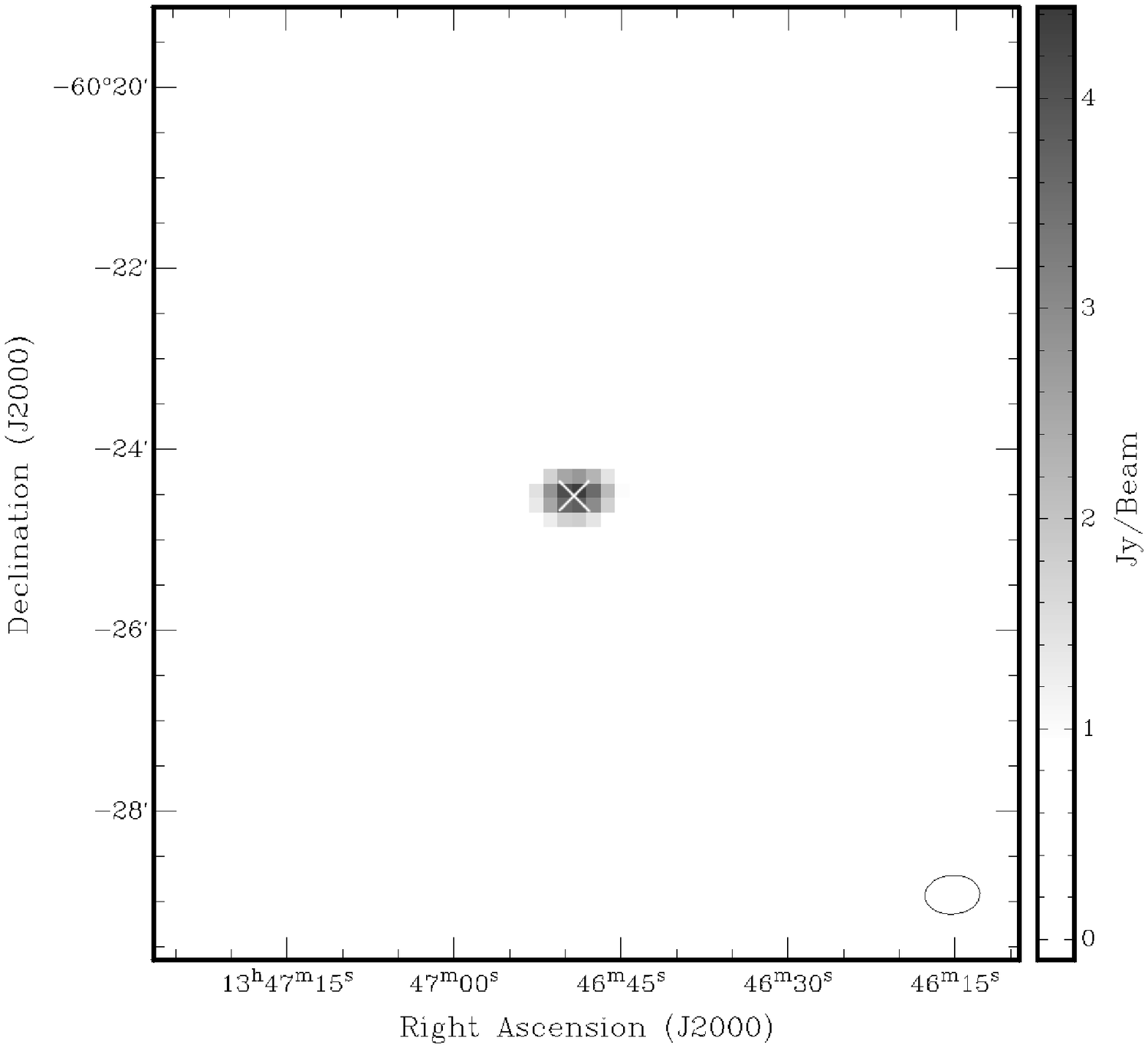}
 \includegraphics[height=0.31\textwidth,angle=270]{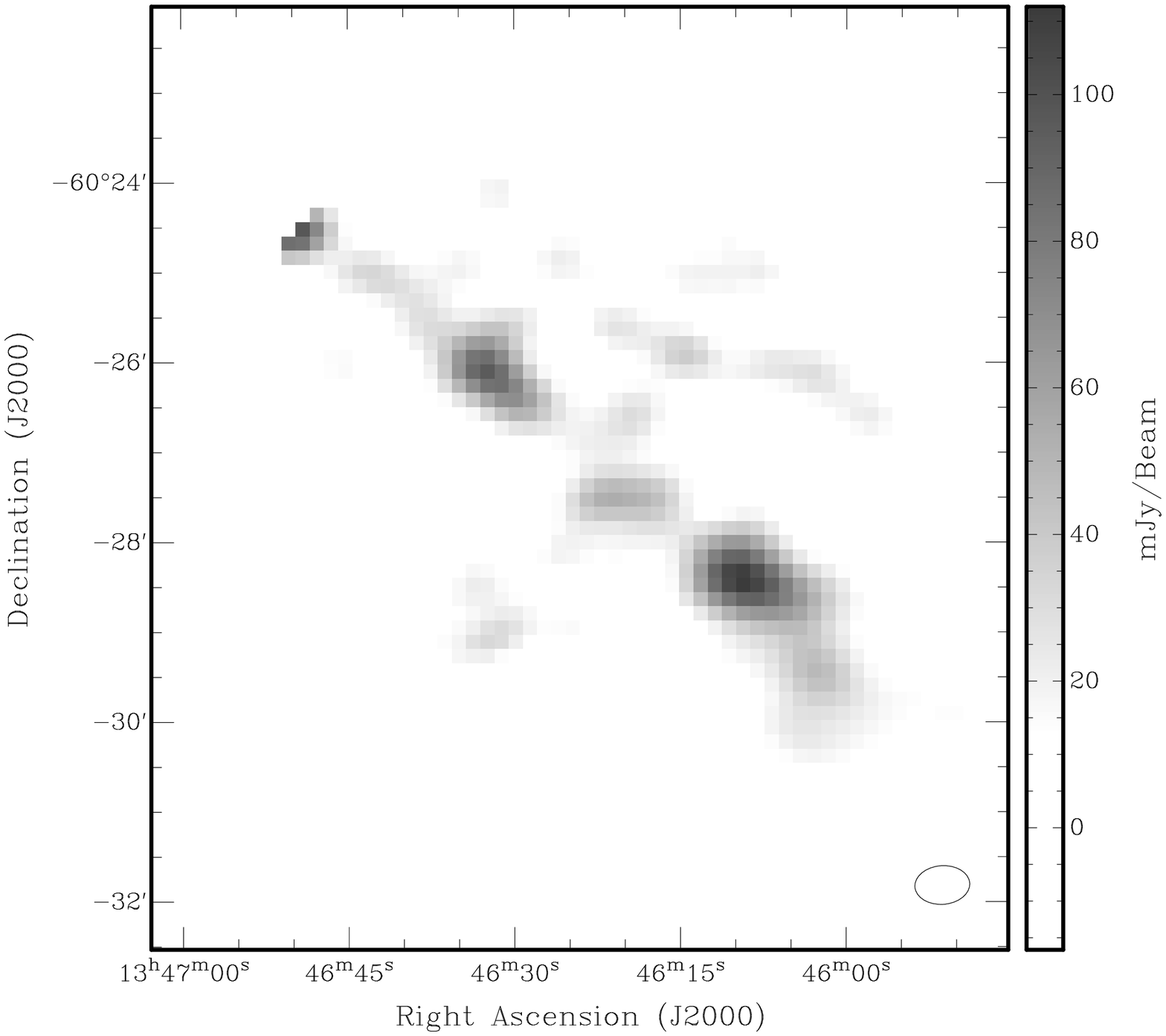}
}

 \subfigure[]
{
 \includegraphics[width=0.31\textwidth]{ps/eastlobe-pol.ps}
 \includegraphics[width=0.31\textwidth]{ps/core-pol.ps}
 \includegraphics[width=0.31\textwidth]{ps/westlobe-pol.ps}
}
\end{figure*}

\begin{figure*}

\caption{\textbf{PKS 1610-60:} \small The beam
  shape is indicated in the lower left hand corner of the images in Panels (a)
  and (b). \textbf{\emph{(a)}} The 18~GHz greyscale image of total intensity.  An "X"
  indicates the position of the associated optical galaxy. \textbf{\emph{(b)}}:
  18~GHz polarisation map; vectors show the \emph{observed} degree and orientation of
  the electric field vectors, while contours trace the total intensity of the
  radio source at levels given in the text below the image.  The rod in the
  lower left hand corner of the image indicates the length of a 100 percent
  polarised component.}\label{fig:1610}
\subfigure[]
{
  \includegraphics[width=1.0\textwidth]{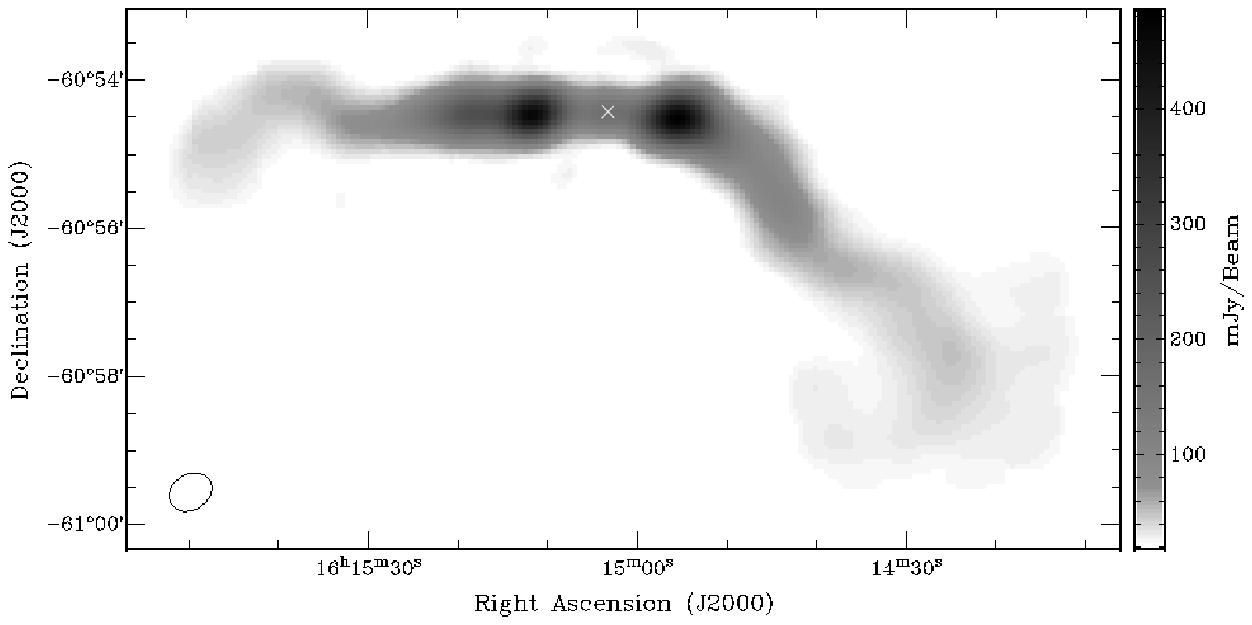}
}

 \subfigure[]
{
 \includegraphics[width=1.0\textwidth]{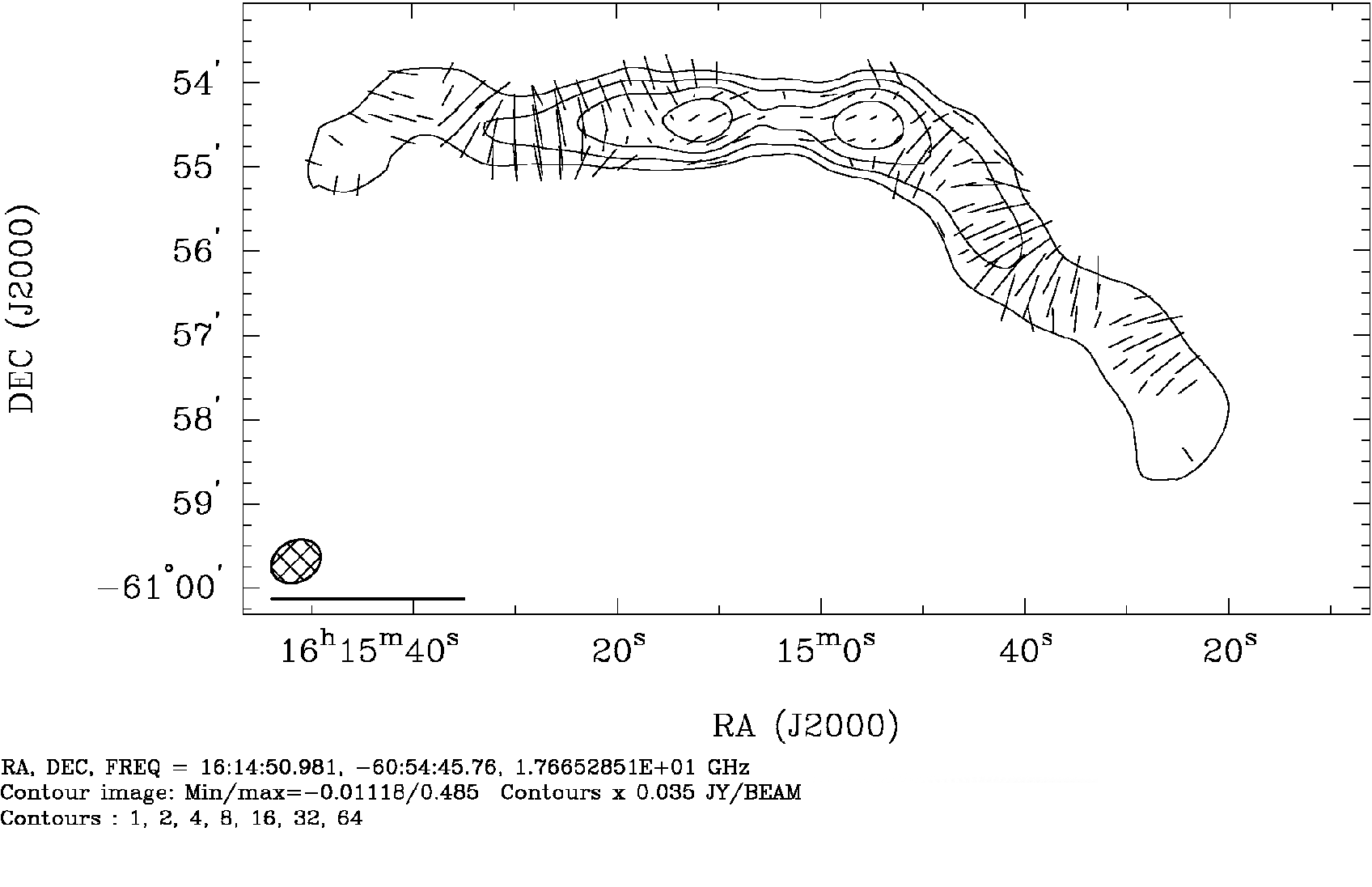}
}
\end{figure*}

\begin{figure*}
\caption{\textbf{PKS 2153-69:} \small The beam
shape is indicated in the lower left hand corner of the
images in Panels (a) and (b). \textbf{\emph{(a)}} The 18~GHz greyscale image of total intensity.  An "X"
indicates the position of the associated optical galaxy. \textbf{\emph{(b)}}
18~GHz polarisation map; vectors show the \emph{observed} degree and
orientation of the electric field vectors, while contours trace the
total intensity of the radio source at levels given in the text below
the image.  The rod in the lower left hand corner of the image
indicates the length of a 100 percent polarised component.}\label{fig:2153}

 \subfigure[]
{
 \includegraphics[width=0.65\textwidth]{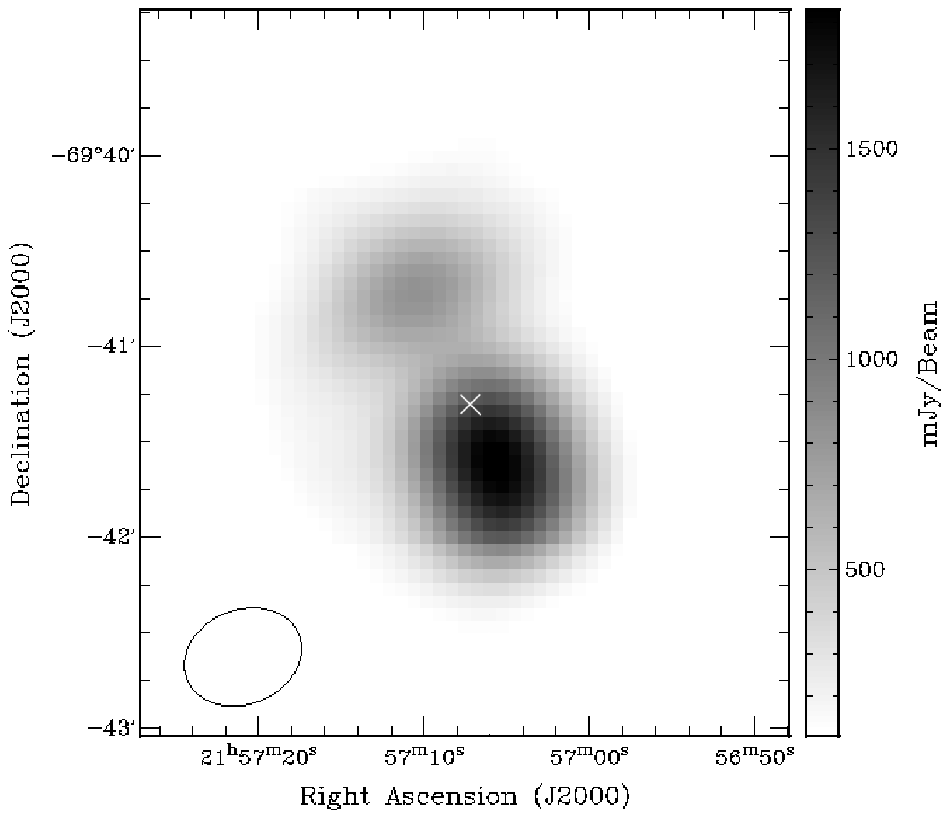}
}

 \subfigure[]
{
 \includegraphics[width=0.55\textwidth]{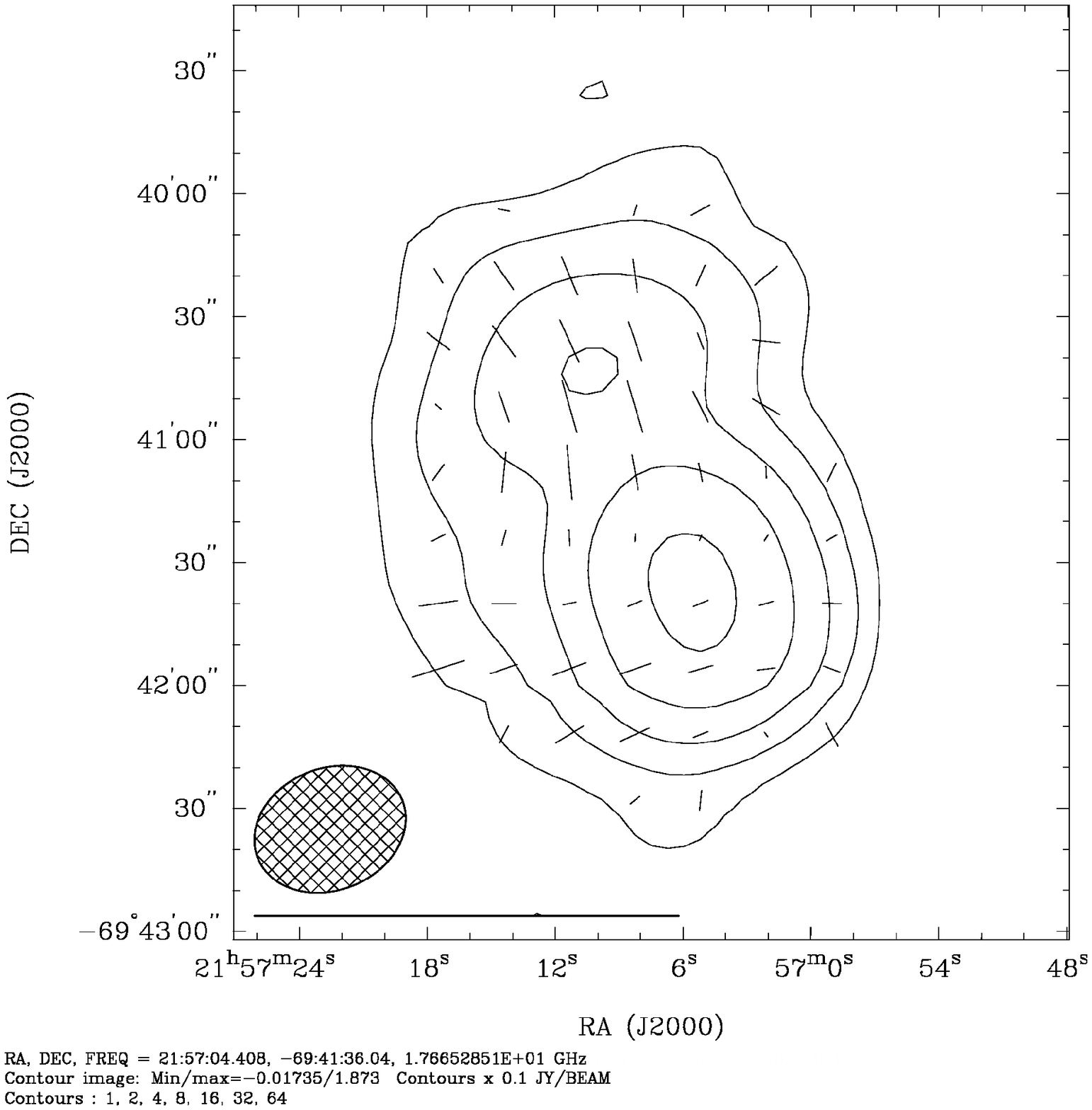}
}
\end{figure*}

\begin{figure*}
\caption{\textbf{PKS 2356-61:} \small The beam
shape is indicated in the lower right hand corner of the
images in Panels (a) and (b). \textbf{\emph{(a)}} The 18~GHz greyscale image of total intensity.  An "X"
indicates the position of the associated optical galaxy. \textbf{\emph{(b)}}
18~GHz polarisation map; vectors show the \emph{observed} degree and
orientation of the electric field vectors, while contours trace the
total intensity of the radio source at levels given in the text below
the image.  The rod in the lower right hand corner of the image
indicates the length of a 100 percent polarised component.}\label{fig:2358}
 \subfigure[]
{
 \includegraphics[width=0.7\textwidth]{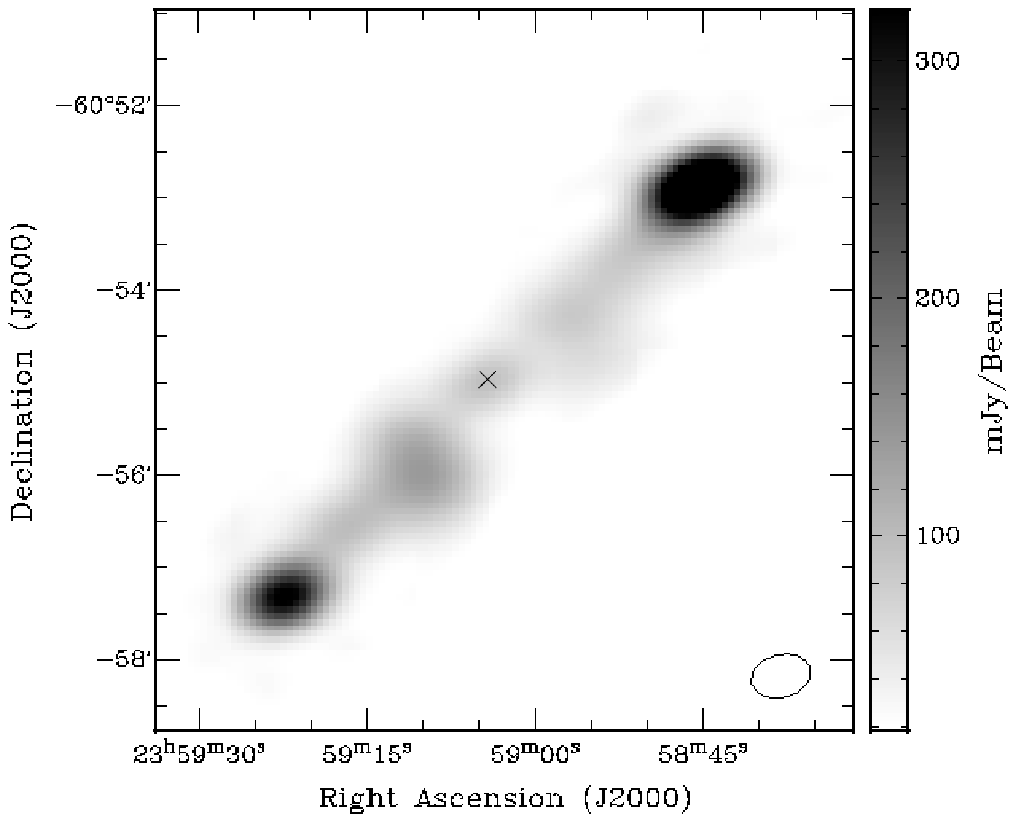}
}

 \subfigure[]
{
 \includegraphics[width=0.6\textwidth]{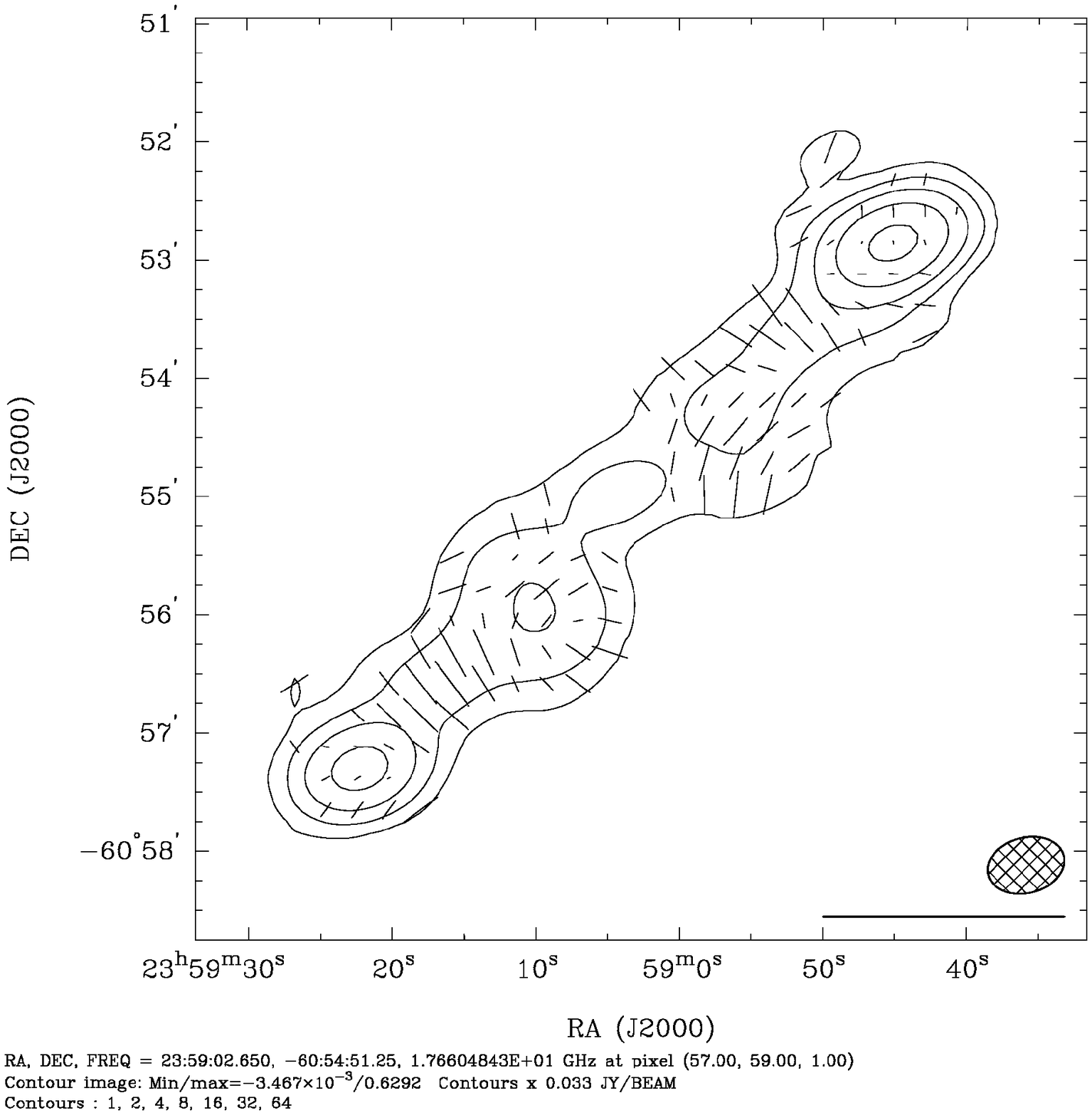}
}
\label{fig:extlast} 
\end{figure*}

The total intensity and fractional polarisation vector maps
whose production are described in the previous section are presented in
Figures \ref{fig:ext}-\ref{fig:extlast}. Greyscale images show the total
intensity of emission with an ``X'' indicating the position of the suspected
host galaxy for each source based on previous identification (as noted in
section \ref{sec:selsum}) with an associated optical galaxy. Sources for which
we did not image the full extent of the low-frequency radio structure are
accompanied by a radio map from previous low-frequency observations to give a
visual reference of the observed subregion. Greyscale images have a linear colour gradient.

The polarisation images indicate the \emph{observed} degree and angle of
polarisation E-field vectors, with contour levels as indicated below the
image. The synthesised beam and a vector showing 100 percent polarisation are
indicated in the lower corner of the image.

\subsection{General Properties at 20\,GHz}
There are several striking and perhaps expected similarities between all the
sources. Six of the objects (apart from PKS 2153-69) have clearly detected
emission from their nuclei, which in some cases were not easily visible in
lower-frequency images; all cores have either undetected or very low values of
polarisation (an upper limit on the polarisation of the cores can be
calculated by taking $3\cdot\sigma_P$ from Table \ref{table:numbers}). The
undetected polarisation of the cores could be the result of several different
effects.
The low values of polarisation 
can be explained by a complex scattering medium surrounding the inner regions
of jet formation, or by a dense plasma. Furthering this supposition are the
high ($\gtrsim 1000~\mbox{rad~m}^{-2}$) rotation measures noted in the inner
regions of some AGN. However, studies of galactic nuclei have been yet unable
to characterise the spectral properties for emission from active galactic
nuclei; the complex spectra are unable to be fit by pure synchrotron self
absorption or free-free spectra. An equally likely case is the beam
depolarisation of close, compact regions of structure that we have not
resolved due to our lower angular resolution. With milliarcsecond
resolution, the pilot VLBA Imaging Polarimetry Survey (VIPS, \citet{VIPS}) has
demonstrated that a sample of 24 active galactic nuclei average below $\sim$3
percent polarisation at 15 GHz, however the full VIPS sample will provide a
more numerically significant assessment for high resolution polarisation.

In several of the sources (esp. in the lobes of Pictor A, Centaurus A,
and the inner hotspots of PKS 1610-60), a notable edge-brightening
effect in fractional polarisation is visible. While this is a feature
common to many extended extragalactic sources, it appears in only a
few of the sources here. The interpretation given by \citet{PRM97} for
this effect and of the perpendicularity of polarisation alignment to
isocontours of total flux density in the western hotspot in Pictor A calls
upon the change in field geometry across the lobe. This interpretation
can be supported by the absence of these effects in the geometry and
strength of observed polarisation in the linear jet regions in several
of the sources here (for instance, the narrow regions of PKS 1610-60
at $\alpha \sim$ 16:15:27 and $\alpha \sim$ 16:14:35-50, and of PKS
2356-61 at $\alpha \sim$ 23:59:17 and $\alpha \sim$ 23:58:53). In
these cases, if there is insignificant or no rotation of polarised
signal, the magnetic field along the jets appears to be aligned with
the jet axis, and across the entire region there is no apparent change
in geometry, nor any apparent limb brightening in polarisation.

\subsection{Individual properties at 20\,GHz}
The inner jets of PKS 1333-33 are unique for this sample in their alignment;
the perpendicular change in magnetic field orientation witnessed in transition
from the core to the jets agrees with that observed by \citet{KBE86}. The
reason for this transition is unclear, though could be explained through
relativistic shock interactions with a medium surrounding the nucleus.
Transverse orientations of magnetic fields are often observed in BL Lacertae
objects, for instance in the VLBI observations of \citet{gabuzda}.

Centaurus B in addition shows a striking alignment fidelity in the outer
regions of the source, with magnetic fields running perpendicularly to the jet
axis along the brighter areas of the lobes. There is a sharp change in
direction and a brightening at the edge of the lobes, and the core is polarised at
approximately 90 degrees to the jet emission. 

In a qualitative way, we can compare the observed electric field
vectors in our images with the lower frequency images of polarised
structure published in previous papers that have been corrected for
rotation measure. Often images are published of magnetic field
orientation, in which case vector images will show vectors
perpendicular to the electric field lines. In the case of Pictor A,
the 5\,GHz images of \citet{PRM97}, which have been corrected for a
rotation measure of 45 \mbox{$rad/m^2$} (which corresponds to a
rotation at 18\,GHz of 0.7 degrees) are in good alignment with our
uncorrected images, where polarisation has been detected by both
observations across all regions of the source. The detailed 20cm
images published by \citet{KBE86} of the inner lobes of PKS 1333-33
show rotation-corrected magnetic fields in the lobes that are
perpendicular to the jet axis. This is again in agreement with the
results of our polarimetry, which show electric field vectors running
directly parallel to the jet axis. To quantify the relative changes in
spectral index and polarisation for different source components in
each source, a more rigourous wide field multifrequency observing
campaign will be necessary.


\begin{figure}
{
 \includegraphics[width=0.35\textwidth,angle=270]{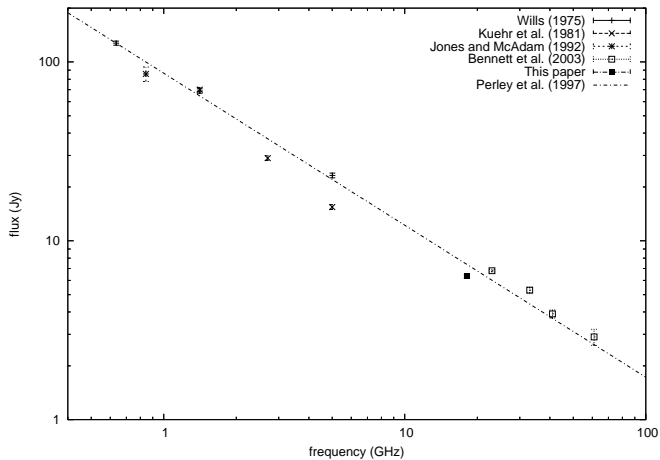}
}
\caption{The spectrum of integrated intensity of Pictor A, including
 measurements from selected literature which have measured its integrated flux
 density. Overplotted is the spectral fit of \citet{PRM97}.}
\label{fig:picaspectrum}
\end{figure}

\subsection{Prospects for CMB mission calibration}
Beam depolarisation imposes a restriction on sources as suitable CMB mission
calibrator candidates. The requirements for a good flux density and
polarisation calibrator for the lower end of Planck's observing frequencies
(the lowest band centre on the LFI is 30\,GHz) include integrated emission $>$
1~Jy across the observation bands, polarised intensity at levels $>$ 200~mJy,
and unresolved emission within the telescope beam. The LFI beam resolutions
range from $\sim$33' at 30\,GHz to 14' at 70\,GHz \citep{PLANCK}. The flight
path and pointing path of the Planck Satellite are such that the north and
south ecliptic poles will be observed often, such that a source in the region
around the south ecliptic pole would be in an ideal position. A limitation on source
size coupled with the depolarisation that will certainly occur within the
large Planck beams means that the brightly polarised northern inner lobe of
Centaurus A, for example, cannot be easily isolated and used for calibration.

The best candidate to arise from this data set is Pictor A, which fits
the Planck requirements in several ways. In total, the source extends
approximately 10 arcminutes; the source yields $\sim0.50$~Jy of
integrated polarised intensity, which is dominated by the brightly
polarised western hotspot. 
Because the emission from the hotspot is highly diffuse and many light years
in linear size, total and polarised emission from this region cannot change rapidly with time, making the source particularly useful for the long
observing term of the Planck mission and more useful than highly polarised
sources which are more compact, variable sources. Conveniently, Pictor A is located within 20
degrees of the south ecliptic pole. The usability of Pictor A as a
high-frequency calibrator is limited by its angular size, which may be marginally resolved at the highest frequency, and its spectral index, which is in
decline at $\nu > 20$\,GHz; both of these could limit the use of Pictor A as a flux
density calibrator at the upper ends of the LFI bands. The integrated spectrum
of Pictor A is plotted in figure \ref{fig:picaspectrum}. Overplotted is the
log-linear spectral fit of \citet{PRM97}, which was fit to data below 5GHz;
the higher frequency WMAP data and the data from our observations generally
follow this log-linear trend, implying that this spectral index will continue
into the relevant Planck bands.
Currently, Pictor A stands as the best candidate from all known extragalactic sources.

\section{Summary}\label{sec:summary}
We have used interferometric wide-field imaging techniques to image
seven of the brightest and most extended sources in the 20~GHz
southern sky region $-30^\circ > \delta$, with the goal of
enhancing the statistical total intensity and polarisation samples
of \citet{BSS} and Burke et al.  (in prep).  The data
analysis included flux density and polarisation measurements and maps
for each of the seven sources.  The images have revealed clearly
detected 20~GHz emission from all galactic nuclei except that of PKS
2153-69.  In all cases, the fractional polarisation of the galactic
core was not detected or less than 1 percent (in the case of Centaurus
A and PKS 1333-33).  While edge-brightening and a perpendicularity of
electric field polarisation vectors to brightness isocontours was
observed in some sources, the lack of such features in compact jet
regions supported the hypothesis that the features may be due to
reorientation of field geometries across broad lobe and hotspot
components.  All sources for which we imaged the entire source had
declining spectra across 5 to 18~GHz at an average spectral index of
-0.88.

PKS 1333-33 was one of the three sources exhibiting detected polarisation in its
core, and it showed a perpendicular field geometry to the close jets spanning
the core. The jets exhibited a transverse magnetic
field configuration to the jet axis, a feature usually observed in BL Lacertae
objects. Centaurus B exhibited a similar change in alignment between its core
and outer regions with very little variation in angle along the peaks of the eastern and
western hotspot regions, however the smaller-scale jets did not have detectable
polarisation. 


We have put forward Pictor A as an excellent extragalactic calibrator
candidate for the ESO Planck Satellite mission that aims to measure the
polarised signature of the CMB. Pictor A's integrated polarised intensity at
0.50$\pm$0.06~Jy, its total intensity of 6.32$\pm$0.11~Jy, its position in the sky and its angular
size contribute to its suitability as a candidate.

\section{Acknowledgements}
R.~D.~Ekers is the recipient of an Australian
Research Council Federation Fellowship, which also provided financial support
for S.~Burke. M.~Massardi wishes to recognize the financial support from ASI
and MUR.  TM acknowledges the support of an ARC Australian Postdoctoral Fellowship (DP0665973).


{\footnotesize

\begin{thebibliography}{}
\bibitem[\protect\citeauthoryear{Baade and Minkowski}{1954}]{BM54}  Baade W., Minkowski R., ApJ, 1954, 119, 215
\bibitem[\protect\citeauthoryear{Bekenstein}{1973}]{Beke73} Bekenstein J.~D., ApJ, 1973, 183, 657
\bibitem[\protect\citeauthoryear{Bennett et al.}{2003}]{WMAP} Bennet C.~L., Hill R.~S., Hinshaw G., Nolta M.~R., Odegard N., Page L., Spergel D.~N., Weiland J.~L., Wright E.~L., Halpern M., Jarosik N., Kogut A., Limon M., Meyer S.~S., Tucker G.~S., Wollack E., ApJS, 2003, 148, 97
\bibitem[\protect\citeauthoryear{Berge and Seielstad}{1967}]{BS67} Berge G.~L. and Seielstad G.~A., ApJ, 1967, 148, 367
\bibitem[\protect\citeauthoryear{Bersanelli and Mandolesi}{2000}]{BM00} Bersanelli M. and Mandolesi N., Astroph. Lett. and Comm., 2000, 37 171
\bibitem[\protect\citeauthoryear{Boehringer et al.}{1996}]{BNS96} Boehringer H., Neumann D.~M., Schindler S., Kraan-Korteweg R.~C., ApJ, 1996, 467, 168
\bibitem[\protect\citeauthoryear{Bolton, Stanley, and Slee}{1949}]{BSS49} Bolton J.~G., Stanley G.~J., and Slee O.~B., Nature, 1949, 164, 101
\bibitem[\protect\citeauthoryear{Bolton, Gardener, and Mackey}{1964}]{BGM64} Bolton J.~G., Gardner F.~F., Mackey M.~B., Aust. J. Phys., 1964, 17, 340
\bibitem[\protect\citeauthoryear{Brown and Burns}{1991}]{BB91} Brown D.~L., Burns J.~O., AJ, 1991, 102, 1917
\bibitem[\protect\citeauthoryear{Burgess and Hunstead}{2006}]{BH06}  Burgess A.~M., Hunstead R.~W., AJ, 2006, 131, 114
\bibitem[\protect\citeauthoryear{Caon, Capaccioli, and D'Onofrio}{1994}]{CCD94} Caon N., Capaccioli M., and D'Onofrio M., AA Supplement, 1994, 106, 199
\bibitem[\protect\citeauthoryear{Christiansen et al.}{1977}]{CFW77} Christiansen W.~N., Frater R.~H., Watkinson A., O'Sullivan J.~D., Lockhart I.~A., Goss W.~M., MNRAS, 1977, 181, 183
\bibitem[\protect\citeauthoryear{Colbert, Mulchaey, and Zabludoff}{2001}]{CMZ01} Colbert J.~W., Mulchaey J.~S., Zabludoff A.~I., AJ, 2001, 121, 808
\bibitem[\protect\citeauthoryear{Condon, Griffith, and Wright}{1993}]{CGW93} Condon J.~J., Griffith M.~R., Wright A.~E., ApJ, 2003, 106, 1095
\bibitem[\protect\citeauthoryear{Cooper, Price, and Cole}{1965}]{coop} Cooper B.~F.~C., Price R.~M., Cole D.~J., AuJPh, 1965, 18, 589
\bibitem[\protect\citeauthoryear{Cornwell}{2008}]{cornwellmultiscale} Cornwell T., 2008, submitted to ``IEEE Special Issue on Signal Processing'', arXiv:0806.2228
\bibitem[\protect\citeauthoryear{Da Costa et al.}{1991}]{DPD91} Da Costa L.~N., Pellegrini P.~S., Davis M., Meiksin A., Sargent W.~L.~W., Tonry J.~L., ApJS, 1991, 75, 935
\bibitem[\protect\citeauthoryear{Danziger, Fosbury, and Penston}{1977}]{DFP77} Danziger I.~J., Fosbury R.~A.~E., Penston M.~V., MNRAS short communication, 1977, 179p, 41
\bibitem[\protect\citeauthoryear{Ekers}{1969}]{Eke69b} Ekers R.~D., Aust. J. Phys. Supplement, 1969, 6
\bibitem[\protect\citeauthoryear{Ekers}{1970}]{Eke69a} Ekers R.~D., Aust. J. Phys., 1970, 23, 217
\bibitem[\protect\citeauthoryear{Ekers et al.}{1983}]{Ekers83}Ekers R.~D., Goss W.~M., Wellington K.~J., Bosma A., Smith R.~M., Schweizer F., AA, 1983, 127, 361
\bibitem[\protect\citeauthoryear{Ekers et al.}{1978}]{Eke1978}Ekers R.~D., Goss W.~M., Kotanyi C.~G., Skellern D.~J., AA, 1978, 69, L21
\bibitem[\protect\citeauthoryear{Emonts et al.}{2008}]{emontsetal} Emonts B.~H.~C., Morganti R., Oosterloo T.~A., Hold J., Tadhunter C.~N., van der Hulst J.~M., Ojha R., Sadler E.~M., MNRAS, 2008, 387, 197
\bibitem[\protect\citeauthoryear{Ferguson}{1989}]{Ferg89} Ferguson H.~C., AJ, 1989, 98, 367
\bibitem[\protect\citeauthoryear{Fosbury et al.}{1998}]{FMW98} Fosbury R.~A.~E., Morganti R., Wilson W., Ekers R.~D., di Serego Alighieri S., Tadhunter  C.~N., MNRAS, 1998, 296, 701
\bibitem[\protect\citeauthoryear{Gabuzda et al.}{2000}]{gabuzda} Gabuzda D.~C., Pushkarev A.~B., Cawthorne T.~V., MNRAS, 2000, 319, 1109
\bibitem[\protect\citeauthoryear{Gardner and Whiteoak}{1966}]{whiteoaketc} Gardner F.~F., Whiteoak J.~B., Ann. Rev. in Astronomy and Astrophysics, 1966, 4, 245
\bibitem[\protect\citeauthoryear{Geldzahler and Fomalont}{1984}]{GF84} Geldzahler B.~J., Fomalont E.~B., AJ, 1984, 89, 1650
\bibitem[\protect\citeauthoryear{Glass}{1979}]{Gla79} Glass I.~S., MNRAS, 1979, 186, 29
\bibitem[\protect\citeauthoryear{Goodlet and Kaiser}{2005}]{GK05} Goodlet, J.~A. and Kaiser, C.~R., MNRAS, 2005, 359, 1456
\bibitem[\protect\citeauthoryear{Goss et al.}{1977}]{GWC77} Goss W.~M., Wellington K.~J., Christiansen W.~N., Watkinson A., Frater R.~H., Little A.~G., Lockhart I.~A., MNRAS, 1977, 178, 525
\bibitem[\protect\citeauthoryear{Graham}{1978}]{Gra78} J.~A. Graham, PASP, 1978, 90, 237
\bibitem[\protect\citeauthoryear{Gregory et al.}{1994}]{GVS94} Gregory P.~C., Vavasour J.~D., Scott W.~K., Condon J.~J., ApJS, 1994, 90, 173 
\bibitem[\protect\citeauthoryear{Haves}{1975}]{Hav75} Haves P., MNRAS, 1975, 173, 553
\bibitem[\protect\citeauthoryear{Herrmann et al.}{2007}]{Herm07} Herrmann F., Hinder I.,  Shoemaker D., Laguna P., Matzner R.~A., ApJ, 2007, 661, 430
\bibitem[\protect\citeauthoryear{Hu, Hedman, and Zaldarriaga}{2003}]{Huetal03} Hu W., Hedman M.M., Zaldarriaga M., Phys. Rev. D, 2003, 67, 4, id. 043004
\bibitem[\protect\citeauthoryear{Huchra and Geller}{1982}]{HG82} Huchra J.~P., Geller M.~J., ApJ, 1982, 257, 423
\bibitem[\protect\citeauthoryear{Jones, Lloyd, and McAdam}{2001}]{JLM01} Jones P.~A., Lloyd B.~D., McAdam W.~B., MNRAS, 2001, 325, 817
\bibitem[\protect\citeauthoryear{Jones and McAdam}{1992}]{JM92} Jones P.~A., McAdam W.~B., ApJS, 1992, 80, 137 
\bibitem[\protect\citeauthoryear{Jones and McAdam}{1996}]{JM96} Jones P.~A., McAdam W.~B., MNRAS, 1996, 282, 137
\bibitem[\protect\citeauthoryear{Killeen, Bicknell, and Ekers}{1986}]{KBE86} Killeen N.~E.~B., Bicknell G.~V., Ekers R.~D., ApJ, 1986, 302, 306
\bibitem[\protect\citeauthoryear{Killeen, Bicknell, and Carter}{1986}]{KBC86} Killeen N.~E.~B., Bicknell G.~V., Carter D., ApJ, 1986, 309, 45
\bibitem[\protect\citeauthoryear{Killeen and Bicknell}{1988}]{KB88}  Killeen N.~E.~B., Bicknell G.~V., ApJ, 1988, 324, 198
\bibitem[\protect\citeauthoryear{Kuehr et al.}{1981}]{Kuehr81} Kuehr H., Witzel A., Pauliny-Toth I.~I.~K., Nauber U., AA Supplement, 1981, 45, 367
\bibitem[\protect\citeauthoryear{Landi, Malizia, and Bassani}{2005}]{LMB05} Landi R., Malizia A., Bassani L., 2005, AA Letters, 441 69L
\bibitem[\protect\citeauthoryear{Laustsen, Schuster, and West}{1977}]{Lau77} Laustsen S., Schuster H.~E., and West R.~M., AA, 1977, 59, L3
\bibitem[\protect\citeauthoryear{Leahy and Fernini}{1989}]{vlamemo161} Leahy P. and Fernini, I., 1989, VLA Scientific Memorandum No. 161
\bibitem[\protect\citeauthoryear{Lewis and Eracleous}{2006}]{LE06} Lewis K.~T., Eracleous M., ApJ, 2006, 624, 711
\bibitem[\protect\citeauthoryear{Loeb}{2007}]{Loeb07} Loeb A., PhRvL, 2007, v99, issue 4, id 041103
\bibitem[\protect\citeauthoryear{Longhetti et al.}{1998}]{Lon98} Longhetti M., Rampazzo R., Bressan A., Chiosi C., AA Supplement, 1998, 130, 267
\bibitem[\protect\citeauthoryear{Lopez et al.}{2007}]{newps} Lopez-Caniego M., Gonzalez-Nuevo J., Herranz D., Massardi M., Sanz J.~L., DeZotti G., Toffolatti L., Argueso F., ApJ Supplement, 2007, 170, 108
\bibitem[\protect\citeauthoryear{Loveday et al.}{1996}]{LPM96} Loveday J., Peterson B.A., Maddoz S.J., Efstathiou G., ApJS, 1996, 107, 201
\bibitem[\protect\citeauthoryear{Lu and Zhou}{2005}]{LZ05} Lu J., Zhou B., ApJ, 2005, 635, L17
\bibitem[\protect\citeauthoryear{Malkan, Gorjian, and Tam}{1998}]{MGT98} Malkan M.~A., Gorjian V., Tam R., ApJS, 1998, 117, 25 
\bibitem[\protect\citeauthoryear{Marenbach and Appenzeller}{1982}]{MA82} Marenbach G., Appenzeller I., AA, 1982, 108, 95
\bibitem[\protect\citeauthoryear{Marshall et al.}{1978}]{MMB78} Marshall F.~E., Mushotzky R.~F., Boldt E.~A., Holt S.~S., Rothschild R.~E., Serlemitsos P.~J., Nature, 1978, 275, 624
\bibitem[\protect\citeauthoryear{Massardi et al.}{2008}]{BSS} Massardi M., Ekers R.~D., Murphy T., Ricci R., Sadler E.~M., Burke S., DeZotti G., Edwards P.~G., Hancock P.~J., Jackson C.~A., Kesteven M.~J., Mahony E., Phillips C.~J., Staveley-Smith L., Subrahmanyan R., Walker M.~A., Wilson W.~E., MNRAS, 2008, 384, 775
\bibitem[\protect\citeauthoryear{Matthews, Morgan, and Schmidt}{1964}]{MMS64} Matthews T.~A., Morgan W.~W., Schmidt M., ApJ, 1964, 140, 35
\bibitem[\protect\citeauthoryear{Mauch et al.}{2003}]{Mau03} Mauch T., Murphy T., Buttery H.J., Curran J., Hunstead R.W., Piestrzynski B., Robertson J.G., Sadler E.M., MNRAS, 2003, 342, 1117
\bibitem[\protect\citeauthoryear{McMullin et al.}{2007}]{CASA} McMullin J.~P., Waters B., Schiebel D., Young W., Golap K.,  Astronomical Data Analysis Software and Systems XIV ASP Conference Series, 2007, Vol. 76, Ed. R. Shaw, F. Hill, D.J. Bell, Astronomical Society of the Pacific, p.127
\bibitem[\protect\citeauthoryear{Michard}{1998}]{Mic98} Michard R., AA, 1998, 334, 453
\bibitem[\protect\citeauthoryear{Mills}{1954}]{Mills54} Mills B.~Y., Observatory, 1954, 74, 248
\bibitem[\protect\citeauthoryear{Perley, Roeser, and Meisenheimer}{1997}]{PRM97} Perley R.~A., Roeser H.~J., Meisenheimer K., AA, 1997, 328, 12
\bibitem[\protect\citeauthoryear{The Planck Consortium}{2005}]{PLANCK} The Planck Scientific Programme, European Space Agency, available from http://www.rssd.esa.int/Planck
\bibitem[\protect\citeauthoryear{Preuss and Fosbury}{1983}]{PF83}  Preuss E. and Fosbury R.~A.~E., MNRAS, 1983, 204, 783
\bibitem[\protect\citeauthoryear{Ricci et al.}{2004}]{Ric04} Ricci R., Sadler E.~M., Ekers R.~D., Stavley-Smith L., Wilson W.~E., Kesteven M.~J., Subrahmanyan R., Walker M.~A., Jackson C.~A., DeZotti G., MNRAS, 2004, 354, 305
\bibitem[\protect\citeauthoryear{Robertson and Roach}{1990}]{RR90} Robertson J.~G., Roach G.~J., MNRAS, 1990, 247, 387
\bibitem[\protect\citeauthoryear{Roeser and Meisenheimer}{1987}]{RM87} Roeser H. and Meisenheimer K., ApJ, 1987, 314, 70
\bibitem[\protect\citeauthoryear{Sadler and Gerhard}{1985}]{SG85} Sadler E.~M., Gerhard O.~E., MNRAS, 1985, 214, 177
\bibitem[\protect\citeauthoryear{Sadler et al.}{2006}]{AT20G2} Sadler E.~M., Ricci R., Ekers R.~D., Ekers J.~A., Hancock P.~J., Jackson C.~A., Kesteven M.~J., Murphy T., Phillips C., Reinfrank R.~F., Staveley-Smith L., Subrahmanyan R., Walker M.~A., Wilson W.~E., de Zotti G., MNRAS, 2006, 371, 898
\bibitem[\protect\citeauthoryear{Sault, Teuben, and Wright}{1993}]{STW95} Sault R.~J., Teuben P~.J., and Wright M.~H.~C, 1995, ASPC, 77, 433
\bibitem[\protect\citeauthoryear{Schilizzi}{1975}]{Sch75} Schilizzi R.~T., Mem. R. astr. Society, 1975, 79, 75
\bibitem[\protect\citeauthoryear{Schilizzi and McAdam}{1975}]{SM75} Schilizzi R.~T., McAdam W.~B., MmRAS, 1975, 79, 1
\bibitem[\protect\citeauthoryear{Schmidt}{1965}]{Sch65} Schmidt M., ApJ, 1965, 141, 1
\bibitem[\protect\citeauthoryear{Schmitt et al.}{2002}]{SPC02} Schmitt H.~R., Pringle J.~E., Clarke C.~J., Kinney A.~L., ApJ, 2002, 575, 150
\bibitem[\protect\citeauthoryear{Simkin et al.}{1999}]{SSS99} Simkin S.~M., Sadler E.~M., Sault R., Tingay S.~J., Callcut J., ApJS, 1999, 123, 447
\bibitem[\protect\citeauthoryear{Smith et al.}{2000}]{SLH00} Smith R.~J., Lucey J.~R., Hudson M.~J., Schlegel D.~J., Davies R.~L., MNRAS, 2000, 313, 469
\bibitem[\protect\citeauthoryear{Subrahmanyan, Saripalli, and Hunstead}{1996}]{SSH96} Subrahmanyan R., Saripalli L., Hunstead R.~W., MNRAS 1996, 279, 257
\bibitem[\protect\citeauthoryear{Tadhunter et al.}{1988}]{TFD88} Tadhunter C.~N., Fosbury R.~A.~E., di Serego Alighieri S., Bland J., Danziger I.~J., Goss W.~M., McAdam W.~B., Snijders M.~A.~J., MNRAS, 1988, 235, 403
\bibitem[\protect\citeauthoryear{Taylor et al.}{2005}]{VIPS} Taylor G.~B., Fassnacht C.~D., Sjouwerman L.~O., Myers S.~T., Ulvestad J.~S., Walker R.~C., Fomalont E.~B., Pearson T.~J., Readhead A.~C.~S., Gehrels N., Michelson P.~F., ApJ Supplement, 2005, 159, 27
\bibitem[\protect\citeauthoryear{Teague, Carter, and Gray}{1990}]{TCG90} Teague P.~F., CarterD. , GrayP.~M. , ApJS, 1990, 72, 715
\bibitem[\protect\citeauthoryear{Thomson, Crane, and Mackay}{1995}]{TCM95} Thomson R.~C., Crane P., Mackay C.~D., ApJ Letters, 1995, 446L, 93
\bibitem[\protect\citeauthoryear{Tingay et al.}{1996}]{TJR96} Tingay S.~J., Jauncey D.~L., Reynolds J.~E., Tzioumis A.K., Migenes V., Gough R., Lovell J.~E.~J., McCulloch P., Costa M.~E., Preston R.~A., Harbison P., AJ, 1996, 111, 718
\bibitem[\protect\citeauthoryear{Tingay et al.}{2000}]{TJR00} Tingay S.~J., Jauncey D.~L., Reynolds J.~E., Tzioumis A.~K., McCulloch P.~M., Ellingsen S.~P., Costa M.~E., Lovell J.~E.~J., Preston R.~A., Simkin S.~M., AJ, 2000, 119, 1695
\bibitem[\protect\citeauthoryear{West and Tarenghi}{1989}]{WT89} West R.~M., Tarenghi M., AA, 1989, 223, 61
\bibitem[\protect\citeauthoryear{Westerlund and Smith}{1966}]{WS66} Westerlund B.~E., Smith L.~F., Aust. J. Phys., 1966, 19, 181
\bibitem[\protect\citeauthoryear{Whiteoak}{1972}]{Whi72} Whiteoak J.~B., Aust. J. Phys., 1972, 25, 233
\bibitem[\protect\citeauthoryear{Wills}{1975}]{wills75} Wills B.~J., Aust. J. Phys. and Astrophysics Supplement, 1975, 38, 1
\bibitem[\protect\citeauthoryear{Wilson, Young, and Shopbell}{2001}]{WYS01} Wilson A.~S., Young A.~J., Shopbell P.~L., ApJ, 2001, 547, 740
\bibitem[\protect\citeauthoryear{Woudt et al.}{2004}]{WKC04} Woudt P.~A., Kraan-Korteweg R.~C., Cayatte V., Balkowski C., Felenbok P., AA, 2004, 415, 9
\bibitem[\protect\citeauthoryear{Wright and Otrupcek}{1990}]{pkscat} Wright A., Otrupcek R., PKS Catalog, 1990
\bibitem[\protect\citeauthoryear{Wright et al.}{1994}]{WGB94} Wright A.~E., Griffith M.~R., Burke B.~F., Ekers R.~D., ApJS, 1994, 91, 111
\bibitem[\protect\citeauthoryear{Wright et al.}{1996}]{WGH96} Wright A.~E., Griffith M.~R., Hunt A.~J., Troup E., Burke B.~F., Ekers R.~D., ApJS, 1996, 103, 145
\bibitem[\protect\citeauthoryear{Young et al.}{2005}]{YWT05} Young A.~J., Wilson A.~S., Tingay S.~J., Heinz S., ApJ, 2005, 622, 830
\bibitem[\protect\citeauthoryear{Younis, Meaburn, and Stewart}{1985}]{YMS85} Younis S., Meaburn J., Stewart P., AA, 1985, 147, 178
\end{thebibliography}

}

\clearpage

\clearpage
\clearpage
\clearpage
\clearpage
\clearpage
\clearpage
\clearpage

\end{document}